\documentclass[12pt]{article}

\usepackage{amsmath}
\usepackage{amssymb}
\usepackage{slashed}
\usepackage[numbers,sort&compress]{natbib}
\usepackage{hyperref}
\usepackage[normalem]{ulem}
\usepackage{cleveref}
\crefname{equation}{Eq.}{Eqs.}
\crefname{figure}{Fig.}{Figs.}
\crefname{table}{Table}{Tables}
\crefname{section}{Section}{Sections}

\newcommand\scalemath[2]{\scalebox{#1} {\mbox{\ensuremath{\displaystyle #2}}}}

\usepackage[letterpaper,margin=1in,bottom=1in]{geometry}
\usepackage{float} % allows forcing plot and table locations with [H]
\usepackage{parskip} % skip spaces between paragraphs
\usepackage{tabulary} % table environment with auto word wrapping
\usepackage{xcolor} % color for tracking edits
\usepackage{soul} % strikeout for tracking edits
\usepackage{subfigure}
\usepackage{graphicx}
\usepackage[section]{placeins} % keep figures inside their sections

\def\lhc2{LHC~Run~II}

\usepackage{cleveref}

\newcommand{\code}[1]{\texttt{#1}}

\def\.4{\vspace{-.5cm}}
\newcommand{\ifb}{~\textrm{fb}^{-1}}

\def\beq{\begin{equation}}
\def\be{\begin{equation}}
\def\beqn{\begin{eqnarray}}
\def\ee{\end{equation}}
\def\eeq{\end{equation}}
\def\eeqn{\end{eqnarray}}

\author{
Amin Aboubrahim\footnote{Email: a.abouibrahim@northeastern.edu}~\ and 
Pran Nath\footnote{Email: p.nath@northeastern.edu}\\~\\
Department of Physics, Northeastern University,
Boston, MA 02115-5000, USA
}

\title{Mixed hidden sector/visible sector dark matter and observation of CP odd Higgs at HL-LHC and HE-LHC}

\begin{document}
\maketitle
%\date

\textbf{Abstract: } 
 
It is very likely that  similar to the case of visible matter, dark matter too  
  is composed of more than one stable component.
 In this work we investigate a two-component dark matter with one component from the visible sector and the other from the 
 hidden sector. Specifically we consider a $U(1)_X$  hidden sector extension of MSSM/SUGRA where we allow for kinetic and Stueckelberg mass
 mixing between the two abelian $U(1)'s$, i.e., $U(1)_X$ and $U(1)_Y$. We further assume that the hidden sector has chiral matter
 which leads to a Dirac fermion as a candidate for dark matter. 
 The lightest neutralino in the visible sector and the Dirac fermion in the hidden sector then constitute the two components of dark matter.
  We investigate in particular MSSM/SUGRA models with radiative breaking occurring on the hyperbolic branch where the Higgs mixing 
  parameter $\mu$ is small (order the electroweak scale) which leads to a lightest neutralino being dominantly a higgsino. While dark matter 
  constituted only of higgsinos  is significantly constrained by data on dark matter relic density and by limits on spin independent
  proton-DM scattering cross section, consistency with data can be achieved
  if only a fraction of the dark matter relic density is constituted of higgsinos with the rest coming from the hidden
  sector.
  An aspect of the proposed model  is the prediction of a relatively light  
  CP odd Higgs $A$ (as well as a CP even $H$  and a charged 
  Higgs $H^{\pm}$) which is observable at HL-LHC and HE-LHC. We perform a detailed collider analysis search for the 
   CP odd Higgs using boosted decision trees in $\tau_h\tau_h$ final states and compare the discovery potential at HL-LHC and HE-LHC. We show that while several of the points among our benchmarks 
 may be observable at HL-LHC, all of them are visible at HE-LHC with much lower integrated luminosities thus reducing significantly the run time for discovery.  Thus the discovery of a CP odd Higgs would lend support to the existence of the hyperbolic branch, a small $\mu$ and
 point to the multi-component nature of dark matter. It is also shown that a part of the parameter space of the extended model
  can  be probed in the next generation direct detection experiments such as XENONnT and LUX-ZEPLIN.

\newpage

\section{Introduction}\label{sec:intro}

The discovery of the Higgs boson at the Large Hadron Collider (LHC) at $\sim 125$ GeV~\cite{Aad:2012tfa,Chatrchyan:2012ufa}, gives strong support for supersymmetry (SUSY). This is so because within the Standard Model (SM) the Higgs boson mass can lie in a wide range up to several hundred GeV in mass~\cite{Lee:1977yc}, while within supersymmetry/supergravity (SUGRA) unified models (for a review see, e.g.~\cite{Nath:2016qzm})
the mass of the Higgs boson is predicted to lie below 130 GeV~\cite{Akula:2011aa,Arbey:2012dq,Akula:2012kk}. In addition to the fact that LHC data respects the supersymmetric limit, stability of the vacuum can be preserved within supersymmetry up the Planck scale while within the standard model the vacuum stability holds only till around $10^{10}$ GeV~\cite{Degrassi:2012ry,Bednyakov:2015sca}. However, as is well known the Higgs boson mass at $\sim 125$ GeV requires a large loop correction within the minimal supersymmetric standard model MSSM/SUGRA which in turn implies that the size of weak scale supersymmetry is large, lying in the several TeV region. This also explains why SUSY has not been  observed at accelerators thus far. The large size of weak scale supersymmetry also has implications for dark matter (DM). In SUGRA unified models, the low energy sparticle spectrum  determined by running the renormalization group equations (RGE) and radiative breaking of the electroweak symmetry  has several branches. On one branch, the ellipsoidal branch, the Higgs mixing  parameter $\mu$ is large and the lightest supersymmetric particle (LSP), the lightest neutralino, is typically a bino. However, in the early universe the binos are not annihilated efficiently 
leading to an LSP relic density far in excess of the current experiment~\cite{Aghanim:2018eyx}. Here one way to  reduce the relic density is through the utilization of coannihilation. Other possibilities within SUGRA models to get conformity with the relic density constraint include a wino-like dark matter or 
a higgsino-like dark matter. 
 
In this work we focus on SUGRA models on the hyperbolic branch with a small $\mu$ (of order the electroweak scale) where the LSP is higgsino-like. Indeed within radiative breaking of the electroweak symmetry higgsino-like dark matter can arise naturally on the hyperbolic branch when $\mu$ is small~\cite{Chan:1997bi,Chattopadhyay:2003xi,Akula:2011jx} (for related works, see, e.g.,~\cite{Feng:1999mn,Baer:2003wx,Feldman:2011ud,Ross:2017kjc}). 
Models of this type are severely constrained by simultaneous satisfaction of dark matter  relic density data and by the 
 spin-independent proton-DM scattering cross section  limits in  direct detection experiments. 
However, such models can be  viable if dark matter is multicomponent with the higgsino-like DM contributing only a fraction of  the relic density with the remainder made up from other sources. Here we discuss a two-component dark matter model where one component is the higgsino (a Majorana fermion) of the visible sector while the
 other component arises from the hidden sector and is a Dirac fermion~\cite{Feldman:2010wy}. 
 Thus the two-component dark matter model is a
 $U(1)_X$ extension of the standard model gauge group where the $U(1)_X$ gauge boson of the hidden sector and the $U(1)_Y$
 gauge boson of the visible sector have both kinetic~\cite{Holdom:1985ag,Holdom:1990xp} 
    and Stueckelberg mass mixings~\cite{Kors:Nath,st-mass-mixing,Feldman:2007wj} (for Stueckelberg extension with an enlarged
    gauge group, see~\cite{Huang:2017bto,Huang:2019obt}).    
      Further, the hidden sector contains matter which provides a Dirac fermion as the second component of dark matter. It is then seen that the Dirac fermion of the hidden sector provides the dominant piece of the relic density but the higgsino dark matter dominates the spin independent cross section in the direct detection experiments. 
 
One remarkable aspect of the two-component model is the prediction of a relatively light CP odd Higgs (in the range of few hundred GeV) which lies in the observable range of the future  generation of colliders (see, e.g.,~\cite{Arkani-Hamed:2015vfh,Mangano:2017tke} and~\cite{Benedikt:2018ofy,Zimmermann:2018koi}). Specifically we focus here on  the high luminosity LHC (HL-LHC) and high energy LHC (HE-LHC). In this work we carry out a detailed
 analysis of the integrated luminosities needed for the observation of this low-lying Higgs. Its observation would lend support to the higgsino nature of the LSP and the multi-component nature of dark matter. At the same time some of the predicted spin-independent scattering cross-sections also lie in the range of the next generation 
 dark matter direct detection experiments. 
The outline of the rest of the paper is as follows: Details of the two-component model are  discussed in section~\ref{sec:model}. 
The scalar sector of the theory is further elaborated in section~\ref{sec:ssector}.
In section~\ref{sec:benchmarks} we give ten representative benchmarks satisfying the relic density 
constraint along with the Higgs boson mass constraint. 
An analysis of the  two-component dark matter, of relic density  and of direct detection is discussed in section~\ref{sec:DM}.
Associated production of CP odd Higgs along with heavy quarks at the LHC is discussed in section~\ref{sec:lhcproduction} followed by the prospects of discovering a CP odd Higgs at HL-LHC and HE-LHC in section~\ref{sec:signature}. Conclusions are given in section~\ref{sec:conc}.
It is also shown that part of the parameter space of the extended model
  can be probed in the next generation direct detection experiments such as XENONnT and LUX-ZEPLIN.
  
We note in passing that there are a variety of supersymmetric $U(1)$ extensions and their effect on DM and collider analyses have been studied extensively in the literature~\cite{U1extensions,Aboubrahim:2019qpc}. We also note that the two-component model can be
easily extended to include other forms of dark matter such as an axion~\cite{Bae:2014rfa}
or an ultralight axion~\cite{Marsh:2015xka,Hui:2016ltb,Halverson:2017deq}.
Further, several works on the HL-LHC and HE-LHC discovery potential have appeared recently and in the past years~\cite{Aboubrahim:2018tpf,Aboubrahim:2018bil,Aboubrahim:2017wjl,Aboubrahim:2017aen,Han:2019grb,Baer:2018hpb}.

\section{The model}\label{sec:model}

As discussed above  we consider an extension  of the standard model gauge group by an additional abelian gauge group $U(1)_X$ of gauge coupling strength $g_X$. The MSSM particle spectrum in the visible sector, i.e., quarks, leptons, Higgs and their superpartners are assumed neutral under $U(1)_X$. Thus the abelian gauge sector of the extended model contains two vector superfields, a vector superfield $B$ associated with the hypercharge gauge group $U(1)_Y$, a vector superfield $C$ associated with the hidden sector gauge group $U(1)_X$, and a chiral scalar superfield $S$. In the Wess-Zumino gauge the $B$ and $C$ superfields have the following components 
\begin{equation}
B=-\theta\sigma^{\mu}\bar{\theta}B_{\mu}+i\theta\theta\bar{\theta}\bar{\lambda}_B-i\bar{\theta}\bar{\theta}\theta\lambda_B+\frac{1}{2}\theta\theta\bar{\theta}\bar{\theta}D_B,
\end{equation}
and
\begin{align}
C=-\theta\sigma^{\mu}\bar{\theta}C_{\mu}+i\theta\theta\bar{\theta}\bar{\lambda}_{C}-i\bar{\theta}\bar{\theta}\theta\lambda_{C}+\frac{1}{2}\theta\theta\bar{\theta}\bar{\theta}D_{C}.
\end{align}
The chiral scalar superfield $S$ can be expanded in terms of its component fields as 
\begin{equation}
\begin{aligned}
S = &\frac{1}{2}(\rho+i a)+\theta\chi+i\theta\sigma^{\mu}\bar{\theta}\frac{1}{2}(\partial_{\mu}\rho+i\partial_{\mu}a) \\ 
&+\theta\theta F+\frac{i}{2}\theta\theta\bar{\theta}\bar{\sigma}^{\mu}\partial_{\mu}\chi+\frac{1}{8}\theta\theta\bar{\theta}\bar{\theta}(\square\rho+i\square a).
\end{aligned}
\end{equation}
The gauge kinetic energy  sector of the model is given by
\begin{equation}
\mathcal{L}_{\rm gk}=-\frac{1}{4}(B_{\mu\nu}B^{\mu\nu}+C_{\mu\nu}C^{\mu\nu})-i\lambda_B\sigma^{\mu}\partial_{\mu}\bar{\lambda}_B-i\lambda_{C}\sigma^{\mu}\partial_{\mu}\bar{\lambda}_{C}+\frac{1}{2}(D^2_B+D^2_{C}).
\label{kinetic-1}
\end{equation} 
Next we allow  gauge kinetic mixing between the $U(1)_X$ and $U(1)_Y$ sectors with terms of the form
\begin{equation}
-\frac{\delta}{2}B^{\mu\nu}C_{\mu\nu}-i\delta(\lambda_{C}\sigma^{\mu}\partial_{\mu}\bar{\lambda}_B+\lambda_{B}\sigma^{\mu}\partial_{\mu}\bar{\lambda}_{C})+\delta D_B D_{C}.
\label{kinetic-2}
\end{equation}
As a result of Eq.~(\ref{kinetic-2}) the hidden sector interacts with the MSSM fields via the small kinetic mixing parameter $\delta$. The kinetic terms in Eq.~(\ref{kinetic-1}) and Eq.~(\ref{kinetic-2})  can be diagonalized by the transformation
\beqn
\left(\begin{matrix} B^{\mu} \cr 
C^{\mu} 
\end{matrix}\right) = \left(\begin{matrix} 1 & -s_{\delta} \cr 
0 & c_{\delta} 
\end{matrix}\right)\left(\begin{matrix} B'^{\mu} \cr 
C'^{\mu} 
\end{matrix}\right), 
\label{rotation}
\eeqn  
where $c_{\delta}=1/(1-\delta^2)^{1/2}$ and $s_{\delta}=\delta/(1-\delta^2)^{1/2}$. 

Aside from gauge kinetic mixing, we assume a Stueckelberg mass mixing between the $U(1)_X$ and $U(1)_Y$ 
sectors so that \cite{Kors:Nath}
\begin{equation}
\mathcal{L}_{\rm St}=\int d\theta^2 d\bar{\theta}^2(M_1 C+M_2 B+S+\bar{S})^2.
\label{lag}
\end{equation}  
We note that Eq.~(\ref{lag}) is invariant under $U(1)_Y$ and $U(1)_X$ gauge transformation so that,
\begin{equation}
\begin{aligned}
&\delta_Y B = \Lambda_Y+\bar{\Lambda}_Y, \, \, \, \, \, \delta_Y S = -M_2\Lambda_Y, \\ 
& \delta_X C = \Lambda_X+\bar{\Lambda}_X, \, \, \, \delta_X S = -M_1\Lambda_X.
\end{aligned}
\end{equation}
In component notation, $\mathcal{L}_{\rm St}$ is 
\begin{equation}
\begin{aligned}
\mathcal{L}_{\rm St} = &-\frac{1}{2}(M_1 C_{\mu}+M_2 B_{\mu}+\partial_{\mu}a)^2-\frac{1}{2}(\partial_{\mu}\rho)^2-i\chi\sigma^{\mu}\partial_{\mu}\bar{\chi}+2|F|^2 \\ 
&+\rho(M_1 D_{C}+M_2 D_B)+\bar{\chi}(M_1\bar{\lambda}_{C}+M_2\bar{\lambda}_B)+\chi(M_1\lambda_{C}+M_2\lambda_B). 
\end{aligned}
\end{equation}
%\gr{The $U(1)_X$ gauge group is not broken by the Higgs mechanism and so,}
In the unitary gauge, the axion field $a$ is absorbed to generate mass for the $U(1)_X$ gauge boson.  \\
The matter sector of the model consists of the visible sector chiral superfields denoted by $\Phi_i$ where $i$ runs over all quarks, squarks, leptons, sleptons, Higgs and Higgsino fields of the MSSM and hidden sector chiral superfields denoted by $\Psi_i$. The Lagrangian for the matter interacting with the $U(1)$ gauge fields is given by
\begin{equation}
\mathcal{L}_{\rm m}=\int d^2\theta d^2\bar{\theta}\sum_i \left[\bar{\Phi}_i e^{2g_Y Y B+2g_X X C}\Phi_i + \bar{\Psi}_i e^{2g_Y Y B+2g_X X C}\Psi_i\right],
\end{equation} 
where $Y$ is the $U(1)_Y$ hypercharge and $X$ is the $U(1)_X$ charge. The MSSM fields are not charged under the hidden sector and vice-versa, i.e., $X\Phi_i=0$ and $Y\Psi_i=0$. The minimal particle content of the hidden sector consists of a left chiral multiplet $\Psi=(\phi, f, F)$ and a charge conjugate $\Psi^c=(\phi', f', F')$ so that $\Psi$ and $\Psi^c$ carry opposite $U(1)_X$ charge and hence constitute an anomaly-free pair. The Dirac field $\psi$ formed by $f$ and $f'$ has a mass $M_{\psi}$ arising from the term $M_{\psi}\Psi\Psi^c$ in the superpotential. Following SUSY breaking, the scalar fields of the hidden sector acquire soft masses equal to $m_0$ (the universal scalar mass of the visible sector) and the additional Dirac mass such that
\begin{equation}
m^2_{\phi}=m^2_0+M^2_{\psi}=m^2_{\phi'}.	
\label{mphi}
\end{equation} 
It is convenient from this point on to introduce Majorana spinors $\psi_S$, $\lambda_X$ and $\lambda_Y$ so that   
 \begin{equation}
  \psi_S =
  \begin{pmatrix}
    \chi_{\alpha}  \\
    \bar{\chi}^{\dot{\alpha}} 
  \end{pmatrix},\quad
  \lambda_X=
  \begin{pmatrix}
    \lambda_{C\alpha}  \\
    \bar{\lambda}^{\dot{\alpha}}_{C} 
  \end{pmatrix},\quad
  \lambda_Y=
  \begin{pmatrix}
    \lambda_{B\alpha}  \\
    \bar{\lambda}^{\dot{\alpha}}_{B}
  \end{pmatrix}.
  \label{spinors}
\end{equation}
In addition to the MSSM soft SUSY breaking terms, we add new terms pertinent to the additional fields
\begin{equation}
\Delta\mathcal{L}_{\rm soft}=-\left(\frac{1}{2}m_X\bar{\lambda}_X\lambda_X+M_{XY}\bar{\lambda}_X\lambda_Y\right)-\frac{1}{2}m^2_{\rho}\rho^2,
\end{equation}
where $m_X$ is the $U(1)_X$ gaugino mass and $M_{XY}$ is the $U(1)_X-U(1)_Y$ mixing mass. \\

After electroweak symmetry breaking,  $\psi_S$ and $\lambda_X$ mix with the MSSM gauginos and higgsinos to form a $6 \times 6$ neutralino mass matrix. We choose as basis $(\lambda_Y$, $\lambda_3$, $\tilde h_1$, $ \tilde h_2$, $\lambda_X$, $\psi_S)$ where the last two fields arise from the extended sector and the first four, i.e., 
 $\lambda_Y, \lambda_3, \tilde h_1, \tilde h_2$ are the gaugino and the
 higgsino fields of the MSSM sector. Using Eq.~(\ref{rotation}) we rotate into the new basis $(\lambda'_Y,\lambda_3,\tilde h_1, \tilde h_2,\lambda'_X,\psi_S)$ so that the $6\times 6$ neutralino mass matrix takes the form
\beqn
\scalemath{0.9}{
\left(\begin{array}{cccc|cc} m_1 & 0 & -c_{\beta}s_W M_Z & s_{\beta}s_W M_Z & -m_1 s_{\delta}+M_{XY}c_{\delta} & M_2  \cr
0 & m_2 & c_{\beta}c_W M_Z & -s_{\beta}c_W M_Z & 0 & 0 \cr
-c_{\beta}s_W M_Z & c_{\beta}c_W M_Z & 0 & -\mu & s_{\delta}c_{\beta}s_W M_Z & 0 \cr
s_{\beta}s_W M_Z & -s_{\beta}c_W M_Z & -\mu & 0 &  -s_{\delta}s_{\beta}s_W M_Z  & 0\cr
\hline
-m_1 s_{\delta}+M_{XY}c_{\delta} & 0 & s_{\delta}c_{\beta}s_W M_Z & -s_{\delta}s_{\beta}s_W M_Z & m_X c^2_{\delta}+m_1 s^2_{\delta}-2M_{XY}c_{\delta}s_{\delta} & M_1 c_{\delta}-M_2 s_{\delta}  \cr
M_2 & 0 & 0 & 0 & M_1 c_{\delta}-M_2 s_{\delta} & 0 \cr     
\end{array}\right)},
\eeqn 
where 
$s_{\beta}\equiv\sin\beta$, $c_{\beta}\equiv\cos\beta$, $s_W\equiv\sin\theta_W$, $c_W\equiv\cos\theta_W$ with $M_Z$ being the $Z$ boson mass. 
 We label the mass eigenstates as 
 \begin{equation}
 \tilde \chi_1^0, ~\tilde \chi_2^0, ~\tilde \chi_3^0, ~\tilde \chi_4^0, ~\tilde \chi_5^0, ~\tilde \chi_6^0\, ,
 \end{equation}
where $\tilde \chi_5^0$ and $\tilde \chi_6^0$ belong to the hidden sector and mix with the usual MSSM neutralinos. In the limit of small mixings between the hidden and the MSSM sectors the masses of the hidden sector neutralinos are 
\begin{equation}
m_{\tilde \chi_5^0}=\sqrt{M_1^2+\frac{1}{4}\tilde m^2_X}-\frac{1}{2}\tilde m_X, \quad \text{and} \quad m_{\tilde \chi_6^0}=\sqrt{M_1^2+\frac{1}{4}\tilde m^2_X}+\frac{1}{2}\tilde m_X.
\end{equation}

We turn now to the charge neutral gauge vector boson sector. Here the $2\times 2$ mass square matrix of the standard model 
is enlarged to become a $3\times 3$ mass square matrix in the $U(1)_X$-extended SUGRA model.
Thus  after spontaneous electroweak symmetry breaking and  the Stueckelberg mass growth the 
$3\times 3$ mass squared matrix of neutral vector bosons in the basis $(C'_{\mu}, B'_{\mu}, A^3_{\mu})$ is given by
\beqn
\mathcal{M}^2_V=\left(\begin{matrix}  M_1^2\kappa^2+\frac{1}{4}g^2_Y v^2 s^2_{\delta} & M_1 M_2\kappa-\frac{1}{4}g^2_Y v^2 s_{\delta} & \frac{1}{4}g_Y g_2 v^2 s_{\delta} \cr
M_1 M_2\kappa-\frac{1}{4}g^2_Y v^2 s_{\delta} & M_2^2+\frac{1}{4}g^2_Y v^2 & -\frac{1}{4}g_Y g_2 v^2 \cr
\frac{1}{4}g_Y g_2 v^2 s_{\delta} & -\frac{1}{4}g_Y g_2 v^2 & \frac{1}{4}g^2_2 v^2 \cr
\end{matrix}\right),
\label{zmassmatrix}
\eeqn
where $A^3_{\mu}$ is the third isospin component, $g_2$ is the $SU(2)_L$ gauge coupling, $\kappa=(c_{\delta}-\epsilon s_{\delta})$, $\epsilon=M_2/M_1$ and $v^2=v^2_u+v^2_d$. The mass-squared matrix of Eq.~(\ref{zmassmatrix}) has one zero eigenvalue which is the photon while the other two eigenvalues are
\begin{align}
&M^2_{\pm} = \frac{1}{2}\Bigg[M_1^2\kappa^2+M^2_2+\frac{1}{4}v^2[g_Y^2 c^2_{\delta}+g_2^2] \nonumber \\
&\pm \sqrt{\left(M_1^2\kappa^2+M^2_2+\frac{1}{4}v^2[g_Y^2 c^2_{\delta}+g_2^2]\right)^2-\Big[M_1^2 g_2^2v^2\kappa^2+M_1^2g^2_Yv^2 c^2_{\delta}+M_2^2g^2_2 v^2\Big]}~\Bigg],
\label{bosons}
\end{align} 
where $M_+$ is identified as the $Z'$ boson mass while $M_-$ as  the $Z$ boson. The diagonalization of the mass-squared matrix of Eq.~(\ref{zmassmatrix}) can be done via two orthogonal transformations where the first is given by~\cite{Feldman:2007wj}
\beqn
\mathcal{O}=\left(\begin{matrix} 1/c_{\delta} & -s_{\delta}/c_{\delta} & 0 \cr
s_{\delta}/c_{\delta} & 1/c_{\delta} & 0 \cr
0 & 0 & 1 \cr
\end{matrix}\right),
\label{omatrix}
\eeqn
which transforms the mass matrix to $\mathcal{M'}^2_V=\mathcal{O}^{T}\mathcal{M}^2_V\mathcal{O}$, 
\beqn
\mathcal{M'}^2_V=\left(\begin{matrix}  M_1^2 & M_1^2\epsilon' & 0 \cr
M_1^2\epsilon' & M_1^2\epsilon'^2+\frac{1}{4}g^2_Y v^2 c^2_{\delta} & -\frac{1}{4}g_Y g_2 v^2 c_{\delta} \cr
0 & -\frac{1}{4}g_Y g_2 v^2 c_{\delta} & \frac{1}{4}g^2_2 v^2 \cr
\end{matrix}\right),
\label{zpmassmatrix}
\eeqn  
where $\epsilon'=\epsilon c_{\delta}-s_{\delta}$.
The gauge eigenstates of $\mathcal{M'}^2_V$ can be rotated into the corresponding mass eigenstates $(Z',Z,\gamma)$ using the second transformation via the rotation matrix
\beqn
\mathcal{R}=\left(\begin{matrix} c_{\eta} c_{\phi}-s_{\theta}s_{\phi}s_\eta & s_\eta c_{\phi}+s_{\theta}s_{\phi}c_\eta & -c_{\theta}s_{\phi} \cr
c_\eta s_{\phi}+s_{\theta}c_{\phi}s_\eta & s_\eta s_{\phi}-s_{\theta}c_{\phi}c_\eta & c_{\theta}c_{\phi} \cr
-c_{\theta} s_\eta & c_{\theta} c_\eta & s_{\theta} \cr
\end{matrix}\right),
\label{rotmatrix}
\eeqn
with $c_{\eta}(c_{\theta})(c_{\phi})\equiv \cos\eta(\cos\theta)(\cos\phi)$ and $s_{\eta}(s_{\theta})(s_{\phi})\equiv \sin\eta(\sin\theta)(\sin\phi)$, where $\eta$ represents the mixing angle between the new gauge sector and the standard model gauge bosons while the other angles are given by
\begin{equation}
\tan\phi=\epsilon', ~~~ \tan\theta=\frac{g_Y}{g_2}c_{\delta}\cos\phi,
\label{angles}
\end{equation}
such that $\mathcal{R}^T\mathcal{M'}^2_V\mathcal{R}=\text{diag}(M^2_{Z'},M^2_{Z},0)$.
The resulting mixing angle is thus given by
\begin{equation}
\tan2\eta\simeq\frac{2\epsilon' M^2_Z\sin\theta}{M^2_{Z'}-M^2_Z+(M^2_{Z'}+M^2_Z-M^2_W)\epsilon'^2},
\end{equation}
with $M_W=g_2 v/2$, $M_{Z'}\equiv M_+$ and $M_{Z}\equiv M_-$.

\section{The scalar sector of the $U(1)_X$-extended MSSM/SUGRA}\label{sec:ssector}

The addition of the chiral scalar superfield $S$ and the hidden sector matter fields bring about new scalar fields to the theory. Thus, the scalar fields of the $U(1)_X$-extended MSSM/SUGRA are the Higgs fields, the scalar $\rho$ and the fields $\phi$ and $\phi'$ of the hidden sector. In the MSSM, the Higgs sector contains two Higgs doublets $H_d$ and $H_u$,
\begin{align}
   H_d &= \begin{pmatrix}
           H^0_d \\
           H^{-}_d \\
         \end{pmatrix}
         ~~\text{and} ~~
    H_u = \begin{pmatrix}
           H^+_u \\
           H^{0}_u \\
         \end{pmatrix},
  \end{align}
with opposite hypercharge which ensures the cancellation of chiral anomalies. Here $H_d$ gives mass to the down-type quarks and 
the leptons while $H_u$ gives mass to up-type quarks. 
The Higgs potential in the MSSM arises from three sources: the $F$ term of the superpotential, the $D$ terms containing the quartic Higgs interaction and the soft SUSY breaking Higgs mass squared, $m_{H_d}^2$ and  $m_{H_u}^2$, and the bilinear $B$ term. The additional scalar field $\rho$ enters the Higgs potential and mixes with the MSSM Higgs doublets. The full CP-conserving Higgs scalar potential in the extended model can be written as
\begin{align}
V_H&=\left[|\mu|^2+m^2_{H_d}-\frac{1}{2}g_Y\rho M_1(\epsilon-s_{\delta})\right]|H_d|^2+\left[|\mu|^2+m^2_{H_u}+\frac{1}{2}g_Y\rho M_1(\epsilon-s_{\delta})\right]|H_u|^2 \nonumber \\ 
&-B\epsilon_{ij}(H^i_u H^j_d + \text{h.c.})+ \left(\frac{g_Y^2c^2_{\delta}+g_2^2}{8}\right)(|H_d|^2-|H_u|^2)^2+\frac{1}{2}g_2^2|H_d^{\dagger}H_u|^2 \nonumber \\
&+ \frac{1}{2}(M^2_1+M^2_2+m^2_{\rho})\rho^2 + \Delta V_{\rm loop} \,,
\label{higgspot}
\end{align}
where $\mu$ is the Higgs mixing parameter appearing in the superpotential term $\mu \hat{H}_u\cdot \hat{H}_d$. The neutral components of the Higgs doublets and the scalar $\rho$ can be expanded  around their VEVs so that
\begin{equation}
\begin{aligned}
H^0_d&=\frac{1}{\sqrt{2}}(v_d+\phi_d + i \psi_d), \\
H^0_u&=\frac{1}{\sqrt{2}}(v_u+ \phi_u  + i \psi_u), \\
\rho&=v_{\rho}+\phi_{\rho}\,.
\label{vev}
\end{aligned}
\end{equation}
The MSSM $2\times 2$ Higgs mass matrix is now extended to become $3\times 3$ with the new scalar field $\phi_{\rho}$ mixing with the two CP even Higgs fields $\phi_d$ and $\phi_u$. As a result, the masses of the CP even Higgses $h$ and $H$ are modified by
amounts proportional to $\epsilon$ and $\delta$ which, however, are small. Similarly, the corrections to the CP odd Higgs mass 
induced by the new sector are negligible. 
 Minimizing the Higgs potential of Eq.~(\ref{higgspot}) in the $\phi_d$, $\phi_u$ and $\phi_{\rho}$ directions we obtain  the constraints
\begin{align}
 m_{H_d}^2+\mu^{2}- B\tan\beta+\dfrac{1}{2}M^2_Z\cos 2\beta+\Delta_{\rm St} &=0, \nonumber \\ 
 m_{H_u}^2 + \mu^{2}-B\cot\beta-\dfrac{1}{2}M^2_Z\cos 2\beta-\Delta_{\rm St} &=0, \nonumber \\ 
 \Big(M_1^2+M_2^2+m_{\rho}^2\Big)v_{\rho}-\frac{1}{4} g_Y v^2 M_1(\epsilon-s_{\delta})\cos 2\beta &=0,
\label{min}
\end{align}
where 
\begin{equation}
\Delta_{\rm St}=-\frac{1}{2} g_Y v_{\rho}M_1(\epsilon-s_{\delta})+\dfrac{1}{8}v^2g_{Y}^2 s^2_{\delta}\cos 2\beta.
\end{equation}
The last of Eqs. (\ref{min}) gives 
\begin{equation}
v_{\rho}=\frac{g_Y v^2\cos 2\beta}{4(M^2_1+M^2_2+m^2_{\rho})}M_1(\epsilon-s_{\delta}),
\end{equation}
which is typically small since $\epsilon$ and $\delta$ are small.

\section{$U(1)_X$-extended MSSM/SUGRA benchmarks}\label{sec:benchmarks}

The particle content of the $U(1)_X$-extended MSSM/SUGRA discussed  in sections~\ref{sec:model} and~\ref{sec:ssector} consists of the particles of the MSSM, {and from the hidden sector
three spin zero particles ($\rho$, $\phi$, $\phi'$),  three spin 1/2 particles (a Dirac fermion $\psi$ and two Majorana neutralinos $\tilde\chi^0_5$, $\tilde\chi^0_6$)  and one massive  vector boson $Z'$.
The model is implemented in the Mathematica package \code{SARAH v4.14.1}~\cite{Staub:2013tta,Staub:2015kfa} which generates model files for \code{SPheno-4.0.3}~\cite{Porod:2003um,Porod:2011nf} which in turn produces the sparticle spectrum and \code{CalcHep/CompHep}~\cite{Pukhov:2004ca,Boos:1994xb} files used by \code{micrOMEGAs-5.0.4}~\cite{Belanger:2014vza} to determine the dark matter relic density and \code{UFO} files which are input to \code{MadGraph5}~\cite{Alwall:2014hca}. 
The input parameters of the $U(1)_X$-extended MSSM/SUGRA~\cite{msugra} with hidden sector matter 
are taken to be 
$m_0, ~~A_0, ~~ m_1, ~~ m_2, ~~ m_3, ~~M_1, ~~m_X, ~~M_{\psi}, ~~B_{\psi}, ~~\delta, ~~g_X, ~~\tan\beta, ~~\text{sgn}(\mu)$,    
where $m_0, ~A_0, ~m_1, ~m_2, ~m_3, ~\tan\beta$ and $\text{sgn}(\mu)$ are the universal scalar mass, the trilinear coupling, the $U(1)$, $SU(2)$ and $SU(3)$ gaugino masses of the MSSM sector and $B_{\psi}$ is the bilinear parameter of the Dirac mass term in the superpotential
all taken to be at the GUT scale. 

Table~\ref{tab1} shows ten representative benchmarks covering a mass range of the CP odd Higgs from $\sim 300$ GeV to 750 GeV. It is to be noted that the larger fraction of the relic density is contributed by the Dirac fermion of the hidden sector while the  higgsino-like neutralinos contribute smaller fraction.

\begin{table}[H]
\begin{center}
\resizebox{1.05\textwidth}{!}{\begin{tabulary}{\textwidth}{l|ccccccccccccc}
\hline\hline\rule{0pt}{3ex}
Model & $m_0$ & $A_0$ & $m_1$ & $m_2$ & $m_3$ & $\mu$ & $M_1$ & $m_X$ & $M_{\psi}$ & $B_{\psi}$ & $\tan\beta$ & $g_X$ & $\delta$ \\
\hline\rule{0pt}{3ex}  
\!\!(a) & 8115 & -7477 & 6785 & 9115 & 4021 & 423 & 1261 & 27 & 627 & 9283 & 6 & 0.06 & 0.02 \\
(b) & 1743 & 898 & 4551 & 2160 & 4084 & 301 & -1086 & 27 & 627 & 5167 & 10 & 0.07 & 0.02 \\
(c) & 1056 & -920 & 1706 & 3417 & 3396 & 243 & 1059 & 89 & 525 & 2846 & 10 & 0.03 & 0.01 \\
(d) & 8424 & -2488 & 6165 & 3544 & 2466 & 330 & -1469 & 473 & 733 & 4680 & 12 & 0.03 & 0.01 \\
(e) & 2011 & -2462 & 3008 & 5030 & 3833 & 598 & 875 & 38 & 425 & 3248 & 9 & 0.06 & 0.06 \\
(f) & 4637 & -4045 & 7004 & 5480 & 2727 & 511 & -1230 & 372 & 613 & 7557 & 15 & 0.04 & 0.04 \\
(g) & 819 & 477 & 7847 & 1218 & 3040 & 201 & 820 & 509 & 401 & 3425 & 12 & 0.05 & 0.09 \\
(h) & 3881 & -2580 & 7449 & 4870 & 4429 & 268 & 850 & 152 & 419 & 9199 & 13 & 0.08 & 0.02 \\
(i) & 1349 & -2722 & 3938 & 4420 & 2558 & 482 & 1292 & 19 & 636 & 4235 & 15 & 0.07 & 0.08 \\
(j) & 2015 & -4435 & 2695 & 5399 & 2470 & 217 & 1343 & 690 & 670 & 4587 & 11 & 0.03 & 0.03 \\ 
\hline
\end{tabulary}}\end{center}
\caption{Input parameters for the benchmarks used in this analysis. Here $M_{XY}=0=B$ at the GUT scale and  $M_2$ is chosen at the GUT scale so that it is nearly vanishing at the electroweak scale. All masses are in GeV.}
\label{tab1}
\end{table}

The CP odd Higgs mass along with the neutralino, chargino, stop, gluino and stau masses are presented in Table~\ref{tab2}. We also show the mass of the lightest CP even Higgs consistent with the observed 125 GeV Higgs within $\pm 2$ GeV error. In some of those benchmarks, the value of $m_0$ is quite small, for instance, point (g) has $m_0 \sim 800$ GeV while the stop mass is $\sim 5$ TeV. The reason is the large value of $m_3$ which via the RGE running generates squark masses in the several TeV range~\cite{Akula:2013ioa}. With heavy gluinos and stops, experimental limits on their masses from ATLAS and CMS can be evaded. Also, the LSP and chargino masses presented in Table~\ref{tab2} have not yet  been ruled out by experiment.  

\begin{table}[H]
\begin{center}
\begin{tabulary}{1.3\textwidth}{l|CCCCCCCCCCC}
\hline\hline\rule{0pt}{3ex}
Model  & $h$ & $\tilde\chi_1^0$ & $\tilde\chi_1^\pm$ & $\tilde{\tau}$ & $\tilde{\chi}^0_5$ & $\tilde t$ & $\tilde g$ & $A$ & $\Omega h^2$ & $(\Omega h^2)_{\chi}$ & $(\Omega h^2)_{\psi}$ \\
\hline\rule{0pt}{3ex} 
\!\!(a) & 123.3 & 455.9 & 457.1 & 8109 & 1245 & 6343 & 8408 & 305.8 & 0.124 & 0.022 & 0.102 \\
(b) & 123.3 & 322.6 & 324.9 & 2115 & 1008 & 5898 & 8195 & 351.8 & 0.101 & 0.012 & 0.089 \\
(c) & 123.1 & 258.9 & 262.6 & 665.6 & 1015 & 4565 & 6855 & 408.9 & 0.116 & 0.009 & 0.107 \\
(d) & 124.0 & 354.8 & 356.4 & 8425 & 1250 & 6573 & 5467 & 450.8 & 0.117 & 0.019 & 0.098 \\
(e) & 123.9 & 639.5 & 642.2 & 1875 & 851.5 & 4943 & 7712 & 504.2 & 0.106 & 0.042 & 0.064 \\
(f) & 124.7 & 544.3 & 545.7 & 4982 & 1055 & 4314 & 5803 & 547.3 & 0.125 & 0.031 & 0.094 \\
(g) & 123.1 & 212.4 & 215.3 & 1906 & 601.8 & 4646 & 6229 & 604.2 & 0.118 & 0.006 & 0.112 \\
(h) & 125.0 & 289.1 & 290.5 & 4426 & 775.5 & 6109 & 8565 & 650.9 & 0.121 & 0.009 & 0.112 \\
(i) & 124.3 & 510.8 & 512.9 & 1627 & 1276 & 3077 & 5292 & 702.7 & 0.118 & 0.028 & 0.090 \\
(j) & 125.0 & 231.5 & 233.7 & 1845 & 1041 & 2335 & 5164 & 750.3 & 0.113 & 0.008 & 0.105 \\
\hline
\end{tabulary}\end{center}
\caption{Display of the SM-like Higgs boson mass, the stau mass,  the relevant electroweak gaugino masses, the CP odd Higgs mass and the relic density for the benchmarks of Table~\ref{tab1} computed at the electroweak scale. All masses are in GeV. }
\label{tab2}
\end{table}

Similar MSSM benchmark scenarios have appeared in~\cite{Bahl:2019ago} where heavy Higgses can be probed at the LHC using an effective field theory approach with a spectrum containing light charginos and neutralinos while the rest of the SUSY particles are heavy.

The particle spectrum of the model contains an extra neutral massive gauge boson, $Z'$. Stringent constraints are set on the mass of the $Z'$~\cite{Tanabashi:2018oca} and most recently by ATLAS~\cite{Aad:2019fac} using 139$\ifb$ of data. In new physics models containing $Z'$ with SM couplings, the mass limit is set at $m_{Z'}>5.1$ TeV. For a model with an extra $U(1)_X$ with a gauge coupling strength $g_X$, the limit can be written as 
\begin{equation}
\frac{m_{Z'}}{g_X}\gtrsim 12 ~\text{TeV}.
\label{zlimit}
\end{equation}
For the benchmarks of Table~\ref{tab1}, the $Z'$ mass obtained from Eq.~(\ref{bosons}) is $\sim M_1$ since $M_2 \sim 0$ and $s_{\delta} \ll 1$. Thus the spectrum contains a $Z'$ with a mass range of $\sim 800$ GeV to $\sim 1500$ GeV. However, since the $U(1)_X$ coupling $g_X<0.1$, the limit of Eq.~(\ref{zlimit}) is satisfied for all the benchmarks. The smallness of $g_X$ also means that the $Z'$ coupling to SM particles is tiny, therefore its production cross-section at $pp$ colliders is suppressed and thus a $Z'$ in the  mass range noted above is 
consistent with the current experimental constraints.
 Further, from Eq.~(\ref{bosons}), the $Z$ boson mass receives a correction due to gauge kinetic and mass mixings. Knowing that $M_2\ll M_1$ and $s_{\delta}\ll 1$, we can write $M^2_-$ as
\begin{equation}
M^2_-\simeq M^2_Z+\frac{\epsilon}{2}g^2_Yv^2 \frac{s_{\delta}}{c_{\delta}}+\frac{1}{4}g^2_2 v^2\left(\frac{\epsilon}{\kappa}\right)^2.
\end{equation}
For the benchmarks, $\epsilon$ takes values in the range $\mathcal{O}(10^{-4})$$-$$\mathcal{O}(10^{-3})$ with $\kappa\sim 1$
and the
 correction to the $Z$ boson mass falls within the current experimental error bars.

\section{Two-component dark matter and its direct detection}\label{sec:DM}

{ As noted earlier one of the constraints on dark matter models is the relic density constraint which according to the PLANCK collaboration~\cite{Aghanim:2018eyx} is given by 
\begin{equation}
(\Omega h^2)_{\rm PLANCK}=0.1198\pm 0.0012.
\label{relic}
\end{equation}  
Since there are two components to dark matter, the total relic density is the sum of the neutralino and the Dirac fermion relic densities, i.e.,
\begin{equation}
(\Omega h^2)_{\rm DM}=(\Omega h^2)_{\chi} + (\Omega h^2)_{\psi}\,.
\label{relic-sum}
\end{equation}
Further, the spin independent DM-proton cross-section that enters in direct detection experiments is given by 
\begin{equation}
\sigma^{\rm SI}_{\text{DM}-p} = R_{\chi} \sigma^{\rm SI}_{p\chi} + R_{\psi} \sigma^{\rm SI}_{p\psi},
\label{cross-section-sum}
\end{equation}
where
\begin{equation}
\mathcal{R}_{\chi}=(\Omega h^2)_{\chi}/(\Omega h^2)_{\rm PLANCK},~~\text{and} ~~\mathcal{R}_{\psi}=(\Omega h^2)_{\psi}/(\Omega h^2)_{\rm PLANCK}.
\end{equation}
Thus one finds that not only the sum of the neutralino and the Dirac fermion relic densities but also their individual contributions
have observable consequences as seen from Eq.~(\ref{cross-section-sum}).

The main processes that enter in the neutralino and Dirac fermion  relic abundance are
\begin{equation}
\begin{aligned}
\chi\chi &\longleftrightarrow \text{SM~SM}, \\
\psi\bar{\psi} &\longleftrightarrow \text{SM~SM}, \\
\psi\bar{\psi} &\longleftrightarrow \chi\chi\,.
\end{aligned}
\label{DMann}
\end{equation}
Note that the process $\chi\bar{\psi}\leftrightarrow \text{SM~SM}$ cannot happen since the only allowed vertex is $\psi\chi\phi$ and $\phi$ does not couple to SM particles.
In order to calculate the relic density in the two-component model, one must solve the Boltzmann equations for the $\chi$ and $\psi$ number densities   
  $n_{\chi}$ and $n_{\psi}$.
Taking  into consideration the processes in Eq.~(\ref{DMann}), the coupled Boltzmann equations for $n_{\chi}$ 
and $n_{\psi}$ are~\cite{Feldman:2010wy}
\begin{equation}
\begin{aligned}
\frac{dn_{\chi}}{dt}&=-3Hn_{\chi}-\langle \sigma v\rangle_{\chi\chi}(n^2_{\chi}-n^2_{\chi,\rm eq})+\frac{1}{2}\langle \sigma v\rangle_{\psi\bar{\psi}\rightarrow\chi\chi}(n^2_{\psi}-n^2_{\psi,\rm eq}), \\
\frac{dn_{\psi}}{dt}&=-3Hn_{\psi}-\frac{1}{2}\langle \sigma v\rangle_{\psi\bar{\psi}}(n^2_{\psi}-n^2_{\psi,\rm eq}),
\end{aligned}
\label{boltzmann}
\end{equation}
where $\langle \sigma v\rangle_{\chi\chi}$ denotes $\langle \sigma v\rangle_{\chi\chi\rightarrow \text{SM~SM}}$, $\langle \sigma v\rangle_{\psi\psi}$ refers to $\psi\bar{\psi}\rightarrow \text{SM~SM},\chi\chi$ and $n_{\rm eq}$ represents the equilibrium number density. The factor of $1/2$ appearing in some of the terms is due to the fact that $\psi$ is a Dirac fermion. With the exception of point (e) in Table~\ref{tab2}, $M_{\psi}>M_{\tilde\chi^0_1}$ and so $\psi$ freezes-out earlier, i.e. at a higher temperature, $T_f$, than $\tilde\chi^0_1$. Notice the small mass gap between $\tilde\chi^0_1$ and $\tilde\chi^{\pm}_1$ resulting in the activation of the coannihilation channel. The solution to Eqs.~(\ref{boltzmann}) does not have a closed form and must be solved numerically. The total relic density can be expressed as
\begin{equation}
\Omega h^2\simeq
\frac{C_{\chi}}{\int_0^{x^{\chi}_f}\langle \sigma v\rangle_{\chi\chi} dx}+\frac{C_{\psi}}{\int_0^{x^{\psi}_f}\langle \sigma v\rangle_{\psi\bar{\psi}} dx},
\end{equation}    
where $x_f=m/T_f$ ($m$ being the mass of the DM particle) and the $C$'s are constants proportional to $g^{*^{-1/2}}M^{-1}_{\rm Pl}$ with $g^*$ the effective number of degrees of freedom at freeze-out for $\chi$ or $\psi$ and $M_{\rm Pl}$ is the Planck mass. 
 In columns 2$-$4 of Table~\ref{tab3} we give the size of the thermally averaged annihilation cross-sections of the processes in Eq.~(\ref{DMann}). The largest cross-section is that of $\chi\chi\leftrightarrow \text{SM~SM}$ since it has weak scale couplings while reactions involving $\psi$ have cross-sections that are two to five orders of magnitude less than those of  $\chi\chi$. The reason is that $\psi\bar{\psi}$ annihilation proceeds through an $s$-channel exchange of $\gamma, Z, Z'$ with couplings proportional to $g_X$, $s_{\delta}$ and $\epsilon$. The smallness of those parameters renders the cross-section tiny.      

\begin{table}[H]
\begin{center}
\begin{tabulary}{1.3\textwidth}{l|CCCCC}
\hline\hline\rule{0pt}{3ex}
Model  & $\langle \sigma v\rangle_{\chi\chi\rightarrow\rm SM ~SM}$ & $\langle \sigma v\rangle_{\psi\psi\rightarrow\rm SM ~ SM}$& $\langle \sigma v\rangle_{\psi\psi\rightarrow \chi \chi}$ & $\mathcal{R}_{\chi}\times \sigma^{\rm SI}_{p\tilde\chi^0_1}$  & $\mathcal{R}_{\psi}\times \sigma^{\rm SI}_{p\psi}$ \\
 & $\times 10^{-26}$ [cm$^3$/s] & $\times 10^{-28}$ [cm$^3$/s] & $\times 10^{-31}$ [cm$^3$/s] & $\times 10^{-47}$ [cm$^2$] & $\times 10^{-50}$ [cm$^2$] \\
\hline\rule{0pt}{3ex} 
\!\!(a) & 4.68 & 5.55 & 2.17 & 1.09 & 0.19 \\
(b) & 9.62 & 0.51 & 34.3 & 6.86 & 903.4 \\
(c) & 13.5 & 0.03 & 2.51 & 5.63 & 73.2 \\
(d) & 7.68 & 141 & 8380 & 2.77 & 30.0 \\
(e) & 2.59 & 1.65 & 0.01 & 10.67 & 2.41 \\
(f) & 3.37 & 698 & 3440 & 2.04 & 1.30 \\
(g) & 19.4 & 1.23 & 0.86 & 5.47 & 9.79 \\
(h) & 11.1 & 0.69 & 5.47 & 0.52 & 32.69 \\
(i) & 3.82 & 3.23 & 0.33 & 3.63 & 0.96 \\
(j) & 16.5 & 18.6 & 1020 & 0.93 & 145.5 \\
\hline
\end{tabulary}\end{center}
\caption{The thermally averaged annihilation cross-sections, $\langle \sigma v\rangle_{\chi\chi\to \rm SM~SM}$,  
 $\langle \sigma v\rangle_{\psi\psi\to \rm SM~SM}$ and  $\langle \sigma v\rangle_{\psi\psi\to \rm \chi\chi}$,  and the 
 spin-independent proton-DM scattering cross-sections  $\mathcal{R}_{\chi}\times \sigma^{\rm SI}_{p\tilde\chi^0_1}$  and $\mathcal{R}_{\psi}\times \sigma^{\rm SI}_{p\psi}$   for the benchmarks of Table~\ref{tab1}.}
\label{tab3}
\end{table}

We turn now to the spin-independent (SI) proton-DM scattering cross-section in direct detection experiments. The main contribution to the proton-neutralino scattering cross-section comes from a $t$-channel exchange of a Higgs boson ($h/H$) while the proton-Dirac fermion scattering cross-section involves a box diagram with the exchange of the scalar $\phi$ from the hidden sector which explains its small value (last column in Table~\ref{tab3}) compared to the proton-neutralino one (fifth column in Table~\ref{tab3}). To get an idea on where the benchmarks
 lie relative to the current experimental sensitivity of direct detection experiments, we plot in Fig.~\ref{fig-dm} the ten benchmarks along with the most recent experimental limits from XENON1T~\cite{Aprile:2018dbl} for the proton-neutralino cross-section
 (the color code for those points has no meaning here).  
We note that some of these benchmarks may be accessible in future experiments such as XENONnT and LUX-ZEPLIN~\cite{Akerib:2018lyp}. Note that for most of the benchmarks $\mathcal{R}_{\psi}\times \sigma^{\rm SI}_{p\psi}$   is very small and lies well below the sensitivity limit of future experiments while others lie even below the neutrino floor.  

\begin{figure}[H]
 \centering
 	\includegraphics[width=0.7\textwidth]{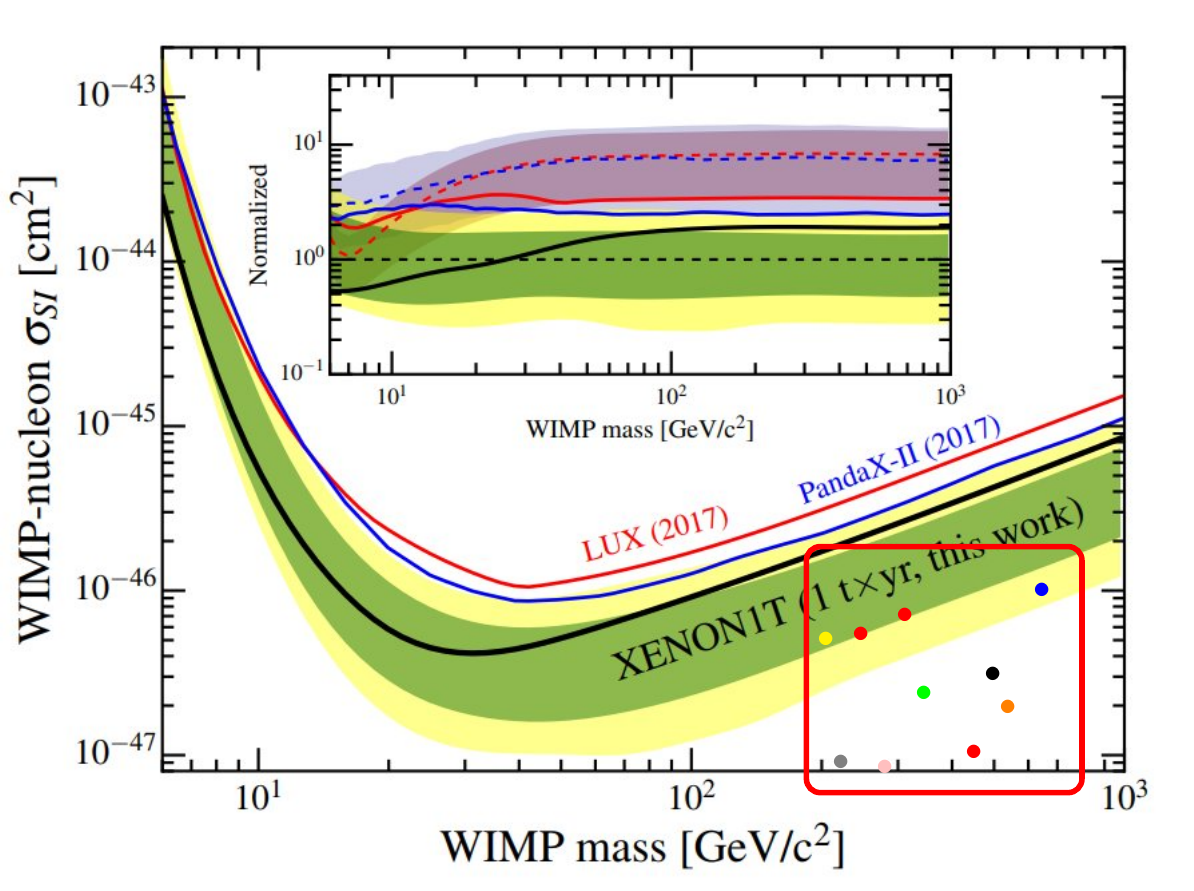}
      \caption{The SI proton-neutralino cross section exclusion limits
as a function of the LSP mass from XENON1T (taken from~\cite{Aprile:2018dbl}). The ten benchmarks are overlaid on the plot showing them lying below the upper limit (black curve). The inset shows the limits from LUX 2017~\cite{Akerib:2016vxi}, PandaX-II~\cite{Cui:2017nnn} and
XENON1T along with the uncertainty bands normalized to the
sensitivity median defined in~\cite{Aprile:2018dbl}.}
	\label{fig-dm}
\end{figure}

}
\section{Associated production of CP odd Higgs  with heavy quarks}
\label{sec:lhcproduction}

As a result of electroweak symmetry breaking and the Stueckelberg mass growth, the Higgs sector of the $U(1)_X$-extended MSSM has six degrees of freedom corresponding to three CP even Higgs, $h, H$ and $\rho$ and one CP odd Higgs $A$ and two charged Higgs $H^{\pm}$. In this section we discuss the production and decay of the CP odd Higgs $A$. Since the weak scale of supersymmetry is high lying in the few TeV region we are in the so-called decoupling limit where the light CP even Higgs $h$ is SM-like and $H$, $A$ and $H^{\pm}$ (charged Higgs) have comparable masses and much greater than $h$. In this case, $A$ exhibits no tree-level couplings to the gauge bosons and couplings to down (up) type fermions that are (inversely) proportional to $\tan\beta$.
For high $\tan\beta$ values, $\tan\beta \gtrsim 10$, the $H, A$ Yukawa couplings to bottom quarks and tau leptons are strongly enhanced, while those to top quarks are strongly suppressed. In this region, the $b$-quark will play an important role as its coupling to the CP odd Higgs is enhanced. For this reason, we examine the associated production of $A$ with bottom-anti bottom quarks, $b\bar{b}A$. 

There are two approaches to calculating the production cross-section of $A$ in association with $b\bar{b}$. The first considers the $b$-quark to be heavy and appearing only in the final state as shown in Fig.~\ref{fig1} where the leading order (LO) partonic processes are
\begin{equation}
gg\rightarrow b\bar{b}A, ~~~ q\bar{q}\rightarrow b\bar{b}A. 
\end{equation}

\begin{figure}[H]
 \centering
 	\includegraphics[width=0.3\textwidth]{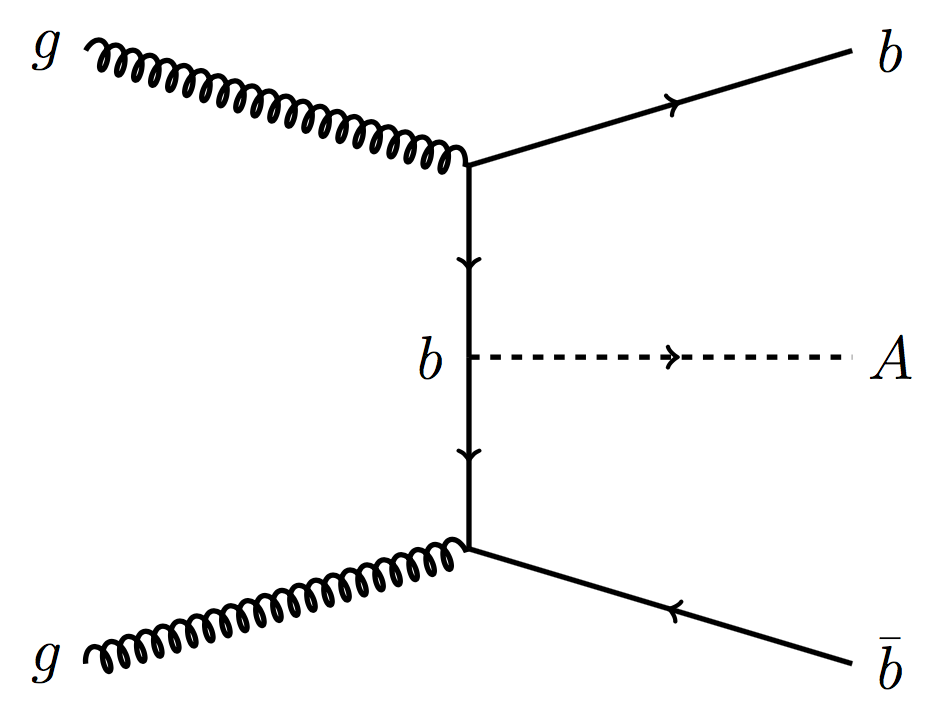} 
 	\hspace{1cm}
 	\includegraphics[width=0.4\textwidth]{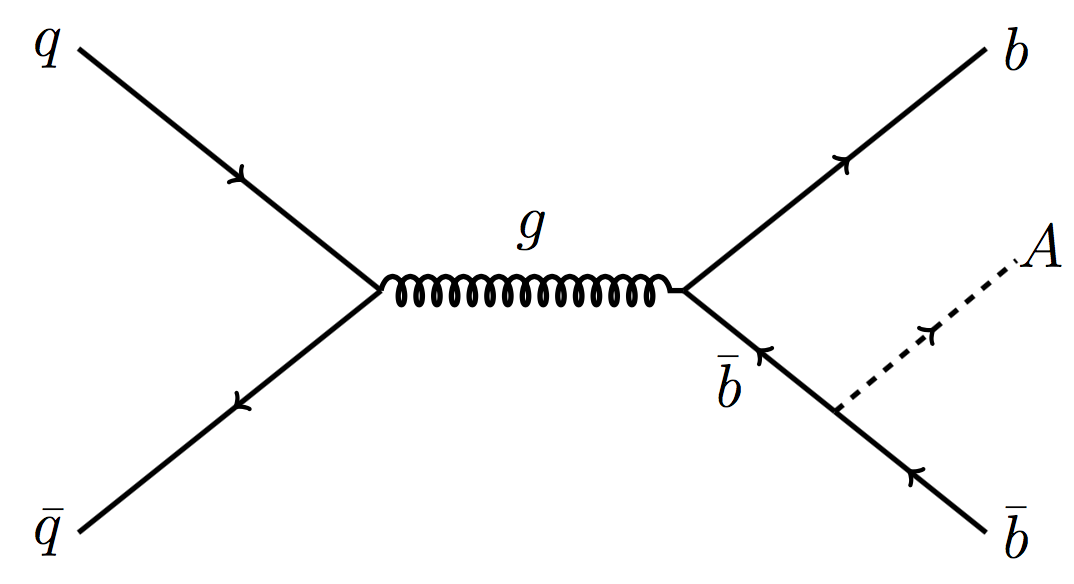}\\
 	\vspace{0.5cm}
 	\includegraphics[width=0.35\textwidth]{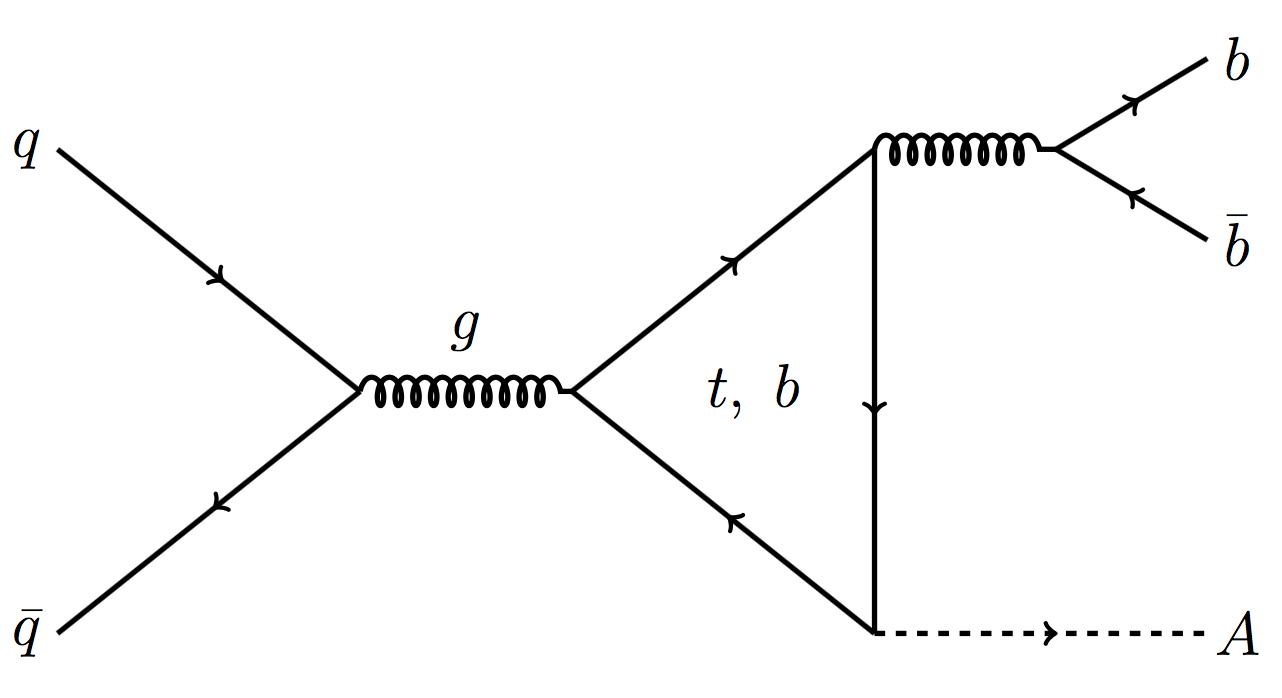}
 	\hspace{0.5cm}
 	\includegraphics[width=0.35\textwidth]{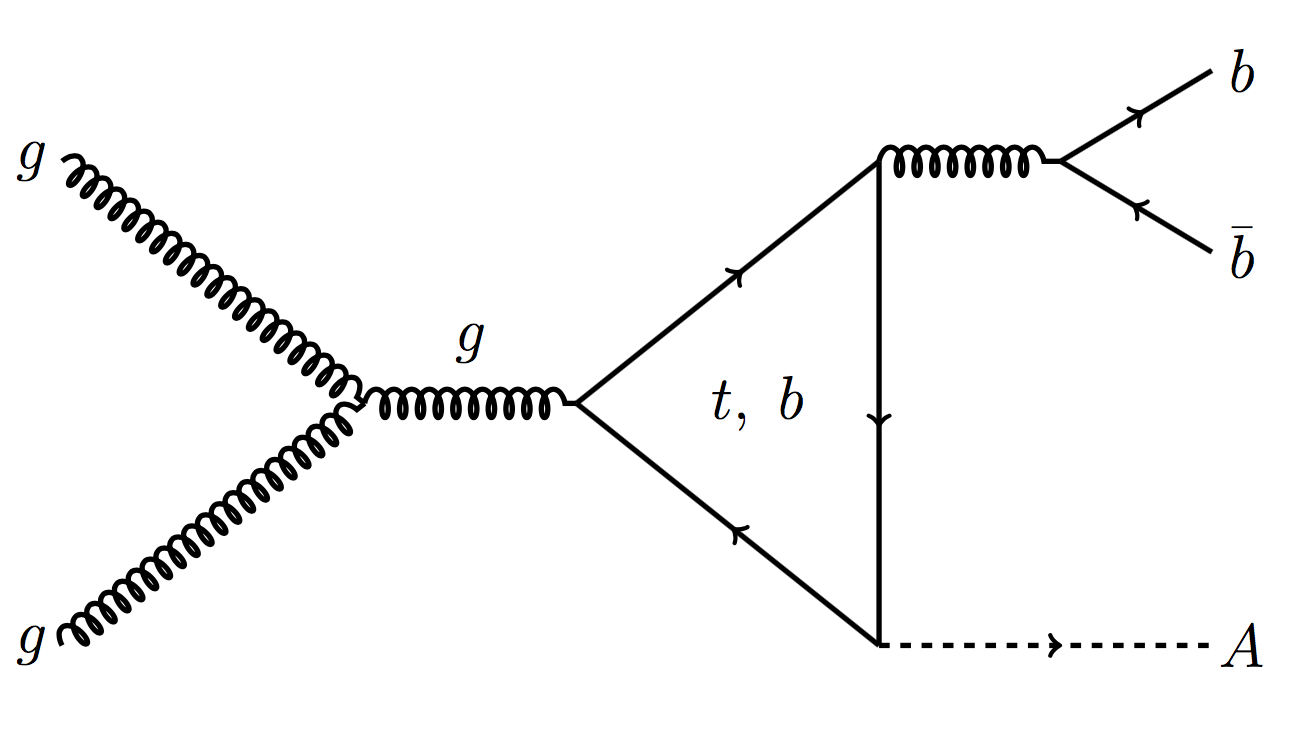}
 	\hspace{0.5cm}
 	\includegraphics[width=0.35\textwidth]{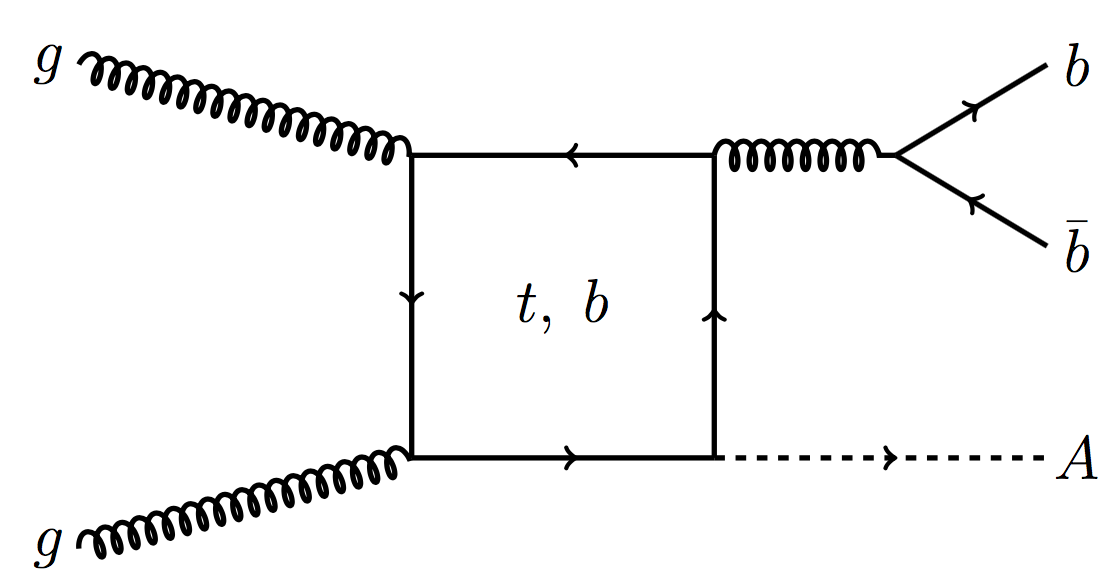}
 	
      \caption{A sample of  the tree (top two) and one-loop (bottom three) Feynman diagrams for $b\bar{b}A$ production at the LHC in the four-flavor scheme. }
	\label{fig1}
\end{figure}

This approach constitutes the four-flavor scheme (4FS) in which the mass of the $b$-quark is considered part of the hard scale of the process. Apart from its dependence on the CP odd Higgs mass and $\tan\beta$, the cross-section is sensitive to the $b$-quark mass which is taken to be the running mass at the appropriate renormalization and factorization scales. The LO $2\rightarrow 3$ diagrams in the 4FS begin at $\mathcal{O}(\alpha^2_S)$ while the next-to-leading order (NLO) diagrams (bottom three diagrams of Fig.~\ref{fig1}) contain bottom and top quarks circulating in the loops. Here the LO cross-section is proportional to $\alpha^2_S y^2_b$ while at NLO, the cross-section is proportional to $\alpha^3_S y^2_b$ and $\alpha^3_S y_b y_t$, where $y_b$, $y_t$ are the bottom and top Yukawa couplings and where the $y_b y_t$ term corresponds to interference between the gluon-fusion (with a top quark in the loop) and $b\bar{b}A$ processes. As mentioned before, the CP odd Higgs coupling to the top quark is suppressed and thus the diagrams involving top quarks do not contribute significantly to the cross-section. \\
Following the prescription of~\cite{deFlorian:2016spz}, the hard scale of the process is chosen at the renormalization and factorization scales such that $\mu_R=\mu_F=(m_A+2m_b)/4$ with $m_A$ the CP odd Higgs mass and $m_b$ being the $b$-quark pole mass while the running $b$-quark mass is $\bar{m}_b(\mu_F)$ (see Table~\ref{tab5}). The 4FS NLO cross-section at fixed order in $\alpha_S$ is calculated with \code{MadGraph5\_aMC@NLO-2.6.3} using \code{FeynRules}~\cite{Alloul:2013bka} \code{UFO} files~\cite{Degrande:2011ua,Degrande:2014vpa} for the Type-II two Higgs doublet model (2HDM). The choice of the latter is justified due to the fact that SUSY-QCD effects for our benchmarks are very minimal since the squarks and gluinos and heavy. The cross-sections at 14 TeV and 27 TeV are displayed in Table~\ref{tab5} along with uncertainties arising from scale variations. 

\begin{table}[H]
\begin{center}
\resizebox{\linewidth}{!}{\begin{tabulary}{1.12\textwidth}{l|CC|CC|CC|CC}
\hline\hline\rule{0pt}{3ex}
Model  & \multicolumn{2}{c}{$\sigma^{\rm 4FS}_{\rm NLO}(pp\rightarrow b\bar{b}A)$} & \multicolumn{2}{c}{$\sigma^{\rm 5FS}_{\rm NNLO}(pp\rightarrow A)$} & \multicolumn{2}{c}{$\sigma^{\rm matched}$} & $\mu_F=\mu_R$ & $\bar{m}_b$\\
  & 14 TeV & 27 TeV & 14 TeV & 27 TeV  & 14 TeV & 27 TeV & \multicolumn{2}{c}{(4FS only)} \\ 
\hline\rule{0pt}{3ex} 
\!\!(a) & 649.4$^{+4.1\%}_{-5.8\%}$ & 2388$^{+2.1\%}_{-5.4\%}$  & 982.0$^{+3.8\%}_{-4.2\%}$  & 3538$^{+4.0\%}_{-4.9\%}$ & 881.0$^{+3.9\%}_{-4.7\%}$  & 3188$^{+3.5\%}_{-5.1\%}$  & 78.5 & 2.91  \\
(b) & 996.9$^{+4.3\%}_{-5.8\%}$ & 3926$^{+1.9\%}_{-5.0\%}$ & 1565$^{+3.5\%}_{-3.6\%}$  & 5963$^{+3.6\%}_{-4.1\%}$  & 1400$^{+3.7\%}_{-4.3\%}$ & 5369$^{+3.1\%}_{-4.4\%}$ & 90.0 & 2.88  \\
(c) & 521.0$^{+4.5\%}_{-6.3\%}$ & 2201$^{+1.9\%}_{-4.6\%}$ & 846.1$^{+3.3\%}_{-3.3\%}$  & 3440$^{+3.4\%}_{-3.7\%}$ & 755.4$^{+3.6\%}_{-4.2\%}$  & 3094$^{+3.7\%}_{-3.2\%}$ & 104.3 & 2.84 \\
(d) & 497.0$^{+5.2\%}_{-7.0\%}$ & 2200$^{+2.9\%}_{-4.7\%}$ & 808.9$^{+3.1\%}_{-3.1\%}$  & 3442$^{+3.2\%}_{-3.3\%}$ & 724.2$^{+3.7\%}_{-4.1\%}$ & 3105$^{+3.1\%}_{-3.7\%}$ & 114.8 & 2.82 \\
(e) & 165.2$^{+5.4\%}_{-7.0\%}$ & 777.8$^{+3.0\%}_{-4.5\%}$ & 277.2$^{+3.0\%}_{-2.9\%}$ & 1247$^{+3.0\%}_{-3.1\%}$ & 247.7$^{+3.6\%}_{-4.0\%}$ & 1123$^{+3.0\%}_{-3.5\%}$ & 128.1 & 2.79 \\
(f) & 313.2$^{+5.0\%}_{-7.5\%}$ & 1524$^{+2.8\%}_{-4.3\%}$ & 530.8$^{+2.9\%}_{-2.8\%}$  & 2493$^{+2.9\%}_{-3.0\%}$ & 474.6$^{+3.4\%}_{-4.0\%}$ & 2243$^{+2.9\%}_{-3.3\%}$ & 138.9 & 2.77  \\
(g) & 125.1$^{+5.1\%}_{-7.8\%}$ & 649.6$^{+3.4\%}_{-4.7\%}$ & 219.6$^{+2.8\%}_{-2.7\%}$ & 1090$^{+2.8\%}_{-2.8\%}$ & 195.9$^{+3.4\%}_{-4.0\%}$ & 979.4$^{+2.9\%}_{-3.3\%}$ & 153.1 & 2.75 \\
(h) & 93.2$^{+5.1\%}_{-8.5\%}$ & 555.0$^{+3.5\%}_{-4.9\%}$ & 182.0$^{+2.7\%}_{-2.6\%}$  & 944.2$^{+2.7\%}_{-2.7\%}$ & 160.1$^{+3.3\%}_{-4.1\%}$ & 848.0$^{+2.90\%}_{-3.3\%}$ & 164.8 & 2.74 \\
(i) & 92.8$^{+5.2\%}_{-8.4\%}$ & 529.9$^{+3.7\%}_{-5.2\%}$ & 164.1$^{+2.6\%}_{-2.6\%}$  & 892.8$^{+2.6\%}_{-2.7\%}$ & 146.8$^{+3.3\%}_{-4.0\%}$ & 804.8$^{+2.9\%}_{-3.3\%}$ & 177.8 & 2.72 \\
(j) & 35.8$^{+5.2\%}_{-8.5\%}$ & 213.5$^{+3.8\%}_{-5.4\%}$ & 65.5$^{+2.5\%}_{-2.6\%}$  & 371.9$^{+2.6\%}_{-2.6\%}$ & 58.4$^{+3.2\%}_{-4.0\%}$ & 334.1$^{+2.9\%}_{-3.3\%}$ & 189.7 & 2.71 \\
\hline
\end{tabulary}}\end{center}
\caption{The production cross-sections, in fb, of the CP odd Higgs in the four flavor scheme at NLO (in association with bottom quarks) and in the five flavor scheme at NNLO along with the matched values at $\sqrt{s}=14$ TeV and $\sqrt{s}=27$ TeV for benchmarks of Table~\ref{tab1}. The running $b$-quark mass, in GeV, is also shown evaluated at the factorization and normalization scales, $\mu_F=\mu_R$ (in GeV).}
\label{tab5}
\end{table}

At any order in perturbation, the 4FS cross-section involves terms $\sim \alpha^n_S\log^n(\mu_F/m_b)$ resulting from collinear splitting of gluons to $b\bar{b}$ pairs. This term is kept under control as long as $\mu_F\sim m_b$, however, this is not the case for our benchmarks especially for larger masses of the CP odd Higgs where such a term spoils perturbative convergence. The way to resolve this issue is by absorbing those terms to all orders in $\alpha_S$. The resummation of those potentially large logarithms is done via the DGLAP evolution of $b$-quark PDFs which constitutes the second approach to calculating cross-sections which is the five-flavor scheme (5FS). In this scheme, $b$-quarks are massless and considered as partons, so they do not appear in the final states at the partonic level. Hence the LO process (zeroth order in $\alpha_S$) in the 5FS for CP odd Higgs production is 
\begin{equation}
b\bar{b}\rightarrow A.
\end{equation}
At the parton level, the 4FS approach has advantage over that of 5FS since realistic $b$-tagging can be done with the former while the latter does not possess this property due to less rich final states. However, the 5FS parton level events are matched to parton showers which add $b$ jets allowing proper $b$-tagging at the analysis level. This is of course pertinent to LO calculations while at higher orders in QCD, the 5FS start to exhibit richer final states with the appearance of $b$-quarks. The 5FS $b\bar{b}A$ production cross-section is known at next-to-NLO (NNLO) and we use \code{SusHi-1.7.0}~\cite{Harlander:2012pb} to determine those cross-sections at 14 TeV and 27 TeV. The renormalization and factorization scales are $\mu_R=m_A$ and $\mu_F=m_A/4$, respectively, which have been shown to be the suitable choices. Scale uncertainties are determined by varying $\mu_R$ and $\mu_F$ such that $\mu_R, 4\mu_F\in\{m_A/2,m_A,2m_A\}$ with $1/2\leq 4\mu_F/\mu_R<2$. Although the $b$-quark is massless, the bottom Yukawa coupling is non-zero and renormalized in the $\overline{\rm MS}$ scheme. The LO cross-section in 5FS is proportional to $y^2_b$ while N$^k$LO is proportional to $y^2_b\alpha^k_S$ with $y_b y_t$ terms vanishing order-by-order in perturbative QCD. In calculating the cross-sections for both 4FS and 5FS cases we have used \code{PDF4LHC15\_nlo\_mc} and \code{PDF4LHC15\_nnlo\_mc}~\cite{Butterworth:2015oua} PDFs, respectively. \\
In order to combine both estimates of the cross-section, we use the Santander matching criterion~\cite{Harlander:2011aa} such that 
\begin{equation}
\sigma^{\rm matched}=\frac{\sigma^{4\rm FS}+\alpha\sigma^{5\rm FS}}{1+\alpha},
\label{matched}
\end{equation}  
where $\alpha=\ln\left(\frac{m_A}{m_b}\right)-2$. The matched cross-section of the inclusive process lies between the 4FS and 5FS values but closer to the 5FS value owing to the weight $\alpha$ which depends on the CP odd Higgs mass. The uncertainties are combined as such, 
\begin{equation}
\delta\sigma^{\rm matched}=\frac{\delta\sigma^{4\rm FS}+\alpha\delta\sigma^{5\rm FS}}{1+\alpha}.
\end{equation}
In Table~\ref{tab5} we give the NLO 4FS, NNLO 5FS and matched cross-sections at 14 TeV and 27 TeV for the ten benchmarks of Table~\ref{tab1} along with $\mu_R$, $\mu_F$ and the running $b$-quark mass in the 4FS case. Notice the dramatic increase in cross-section in going from 14 TeV to 27 TeV due to the production of strongly interacting particles along with $A$. The cross-sections have been checked with publically available results~\cite{deFlorian:2016spz} by a proper scaling of the bottom Yukawa coupling. In the MSSM, the tree-level Higgs Yukawa coupling to bottom quarks is given by
\begin{equation}
y_{bbA}=\frac{\sqrt{2}m_b}{v}\tan\beta,
\end{equation}
for $\tan\beta\gg 1$, where $v$ is the SM VEV. Besides QCD corrections, this Yukawa coupling receives SUSY-QCD corrections given, at one-loop level, by~\cite{Baglio:2010ae}
\begin{equation}
\Delta_b\approx \frac{2\alpha_S}{3\pi}\frac{\mu m_{\tilde g}}{\text{max}(m^2_{\tilde g}, m^2_{\tilde b_1}, m^2_{\tilde b_2})}\tan\beta,
\end{equation}
where $m_{\tilde b_{1,2}}$ and $m_{\tilde g}$  are the sbottom and the gluino masses. Taking this correction into consideration, one then needs to scale the SM $b\bar{b}h$ cross-section by the square of
\begin{equation}
y_{bbA}/y_{bbh}\sim \frac{\tan\beta}{1+\Delta_b},
\end{equation}
in order to obtain the MSSM cross-section. However,  $\Delta_b$ is negligible for our benchmarks due to heavy gluinos and sbottoms. In this case, the scaling only requires multiplying by $\tan^2\beta$ to which we have found reasonable agreement with our results. 

Due to its enhanced coupling to bottom quarks, the CP odd Higgs preferentially decays to $b\bar{b}$ pair while the second largest branching ratio is to $\tau^+\tau^-$ pair as shown in Table~\ref{tab6}. In the MSSM, the branching ratio to $Z~h$ is quite small and is not considered as a significant channel for discovery, at least at 14 TeV.   

\begin{table}[H]
\begin{center}
\begin{tabulary}{1.5\textwidth}{l|CCC}
\hline\hline\rule{0pt}{3ex}
Model  & BR$(A\rightarrow b\bar{b})$ & BR$(A\rightarrow \tau^+\tau^-)$ & BR$(A\rightarrow Z~h)$   \\
\hline\rule{0pt}{3ex} 
\!\!(a) & 0.844 & 0.113 & 0.041 \\
(b) & 0.779 & 0.106 & 0.005 \\
(c) & 0.714 & 0.099 & 0.004 \\
(d) & 0.780 & 0.110 & 0.002 \\
(e) & 0.617 & 0.088 & 0.004 \\
(f) & 0.827 & 0.119 & 0.001 \\
(g) & 0.726 & 0.106 & 0.001 \\
(h) & 0.787 & 0.116 & 0.001  \\
(i) & 0.820 & 0.123 & $<0.001$  \\
(j) & 0.716 & 0.108 & 0.001 \\
\hline
\end{tabulary}\end{center}
\caption{The branching ratios of the CP odd Higgs into standard model particles for the benchmarks of Table~\ref{tab1}. }
\label{tab6}
\end{table}

\section{CP odd Higgs signature in $\tau_h\tau_h$ final state at the LHC}\label{sec:signature}

We begin this section by a review of the experimental status of the MSSM CP odd Higgs. The most recent constraints on the CP odd Higgs mass come from Run 2 results from ATLAS~\cite{Aaboud:2017sjh} and CMS~\cite{Sirunyan:2018zut} collaborations using 36$\ifb$ of data. ATLAS used low scale benchmarks from~\cite{Heinemeyer:2013tqa} satisfying the light Higgs boson mass constraint and characterized by small $\mu$ and SUSY breaking scale (recommended by the LHC-HXSWG). Interpreted in the MSSM ($m_{\rm h}^{\rm mod+}$ model\footnote{Here the top-squark mixing is  fixed so that the lightest CP-even Higgs mass approximates the measured mass~\cite{Carena:2013ytb}}), the results exclude $\tan\beta>5$ for $m_A=250$ GeV and $\tan\beta>51$ for $m_A=1500$ GeV. Due to the absence of light neutralinos in the spectrum, the hMSSM\footnote{Here the measured value of the Higgs boson $h$ is used to predict under certain assumptions the masses 
 and couplings of the MSSM Higgs bosons~\cite{Djouadi:2013uqa,Bagnaschi:2015hka}} provides more stringent constraints because of higher $A\rightarrow \tau\tau$ branching ratio. Thus here $\tan\beta>1$ for $m_A=250$ GeV and $\tan\beta>42$ for $m_A=1500$ GeV are excluded. CMS uses the same benchmarks, cross-sections and branching ratios of the CP odd Higgs and arrives at similar exclusion limits where $m_A\lesssim 250$ GeV is excluded for $\tan\beta>6$ and the exclusion contour reaches 1600 GeV for $\tan\beta=60$. Projections for HL-LHC studies regarding the mass reach for the CP odd Higgs in case no excess is found is available~\cite{Atlas:2019qfx}. The benchmarks of Table~\ref{tab1} are not yet excluded by experiment and lie within the contour set for HL-LHC. It is worth stressing the fact that such interpretations of experimental results as mentioned above are carried out within models that are very different from the one we consider here.

The signal we investigate consists of a CP odd Higgs decaying to two hadronic taus and produced alongside two $b$-quarks which can be tagged. Even in the 5FS, $b$-flavored jets can appear at the parton shower level and so $b$-tagging is viable here too. In order to account for misidentified $b$-tagged jets, we require that our final states contain at least one $b$-tagged jet and two tau-tagged $(\tau_h)$ jets such that $p_T(b)>20$ GeV, $|\eta(b)|<2.5$ and $p_T(\tau_h)>15$ GeV. 
 
The standard model backgrounds relevant to the final states considered here are $t\bar{t}$, $t$+jets, $t+W/Z$, QCD multijet, diboson and $W,Z/\gamma^*$+jets. The signal and SM backgrounds are simulated at LO with \code{MadGraph5-2.6.3} interfaced with LHAPDF~\cite{Buckley:2014ana} and using the \code{NNPDF30LO} PDF set. The cross-sections of the SM backgrounds are then normalized to their NLO values while those of the signal are scaled to their matched values in Table~\ref{tab5}. The parton level events are passed to \code{PYTHIA8}~\cite{Sjostrand:2014zea} for showering and hadronization. A five-flavor MLM matching~\cite{Mangano:2006rw} is performed on the backgrounds to avoid double counting of jets at the shower level. Jets are clustered with \code{FASTJET}~\cite{Cacciari:2011ma} using the anti-$k_t$ algorithm~\cite{Cacciari:2008gp} with jet radius 0.4. Detector simulation and event reconstruction is handled by \code{DELPHES-3.4.2}~\cite{deFavereau:2013fsa} using the new cards for HL-LHC and HE-LHC generic detectors. The resulting files are read and analyzed with \code{ROOT-6.16}~\cite{Antcheva:2011zz}.

Due to the smallness of the signal cross-section in comparison to the SM backgrounds (especially following the selection criteria), we use Boosted Decision Trees (BDT) to separate the signal from the background. The type of BDT used here is known as ``Adaptive BDT" (\code{AdaBoost}). Before giving a brief overview of BDTs, we list the kinematic variables used to help in discriminating the signal from the background:   
\begin{enumerate}
\item The total transverse mass of the di-tau system is given by~\cite{ATLAS:2016fpj}
\begin{equation}
m^{\rm tot}_T=\sqrt{m^2_T(E^{\rm miss}_T,\tau_{h1})+m^2_T(E^{\rm miss}_T,\tau_{h2})+m^2_T(\tau_{h1},\tau_{h2})},
\end{equation}
where
\begin{equation}
m_T(i,j)=\sqrt{2p_T^i p_T^j(1-\cos\Delta\phi_{ij})}.
\end{equation}
This variable has the best separating power especially for heavier CP odd Higgs mass.

\item The hadronic di-tau invariant mass, $m_{\tau_h\tau_h}$, has the same effect as $m^{\rm tot}_T$ and in addition works well for low mass signals.

\item The angular separation $\Delta\phi(\tau_{h1},\tau_{h2})$ between the leading and sub-leading hadronic tau jets. For the signal, the variable is mostly peaked at $\Delta\phi(\tau_{h1},\tau_{h2})>2.8$ while it peaks for small values (near zero) for the background. 

\item The number of charged tracks associated with the leading tau, $N_{\rm tracks}^{\tau}$. Due to its one and three-prong decays, a tau can be identified by the tracks' charge multiplicities. 

\item Due to the presence of $b$-tagged jets, we use the number of such jets, $N_{\rm jet}^b$, as a discriminating variable. 

\item Due to the rich jetty final states, we define the variable $\ln(p_T^{\rm jet})$ as
\begin{equation}
\ln(p_T^{\rm jet})= \begin{cases} 
      \ln(p_T^{\rm jet_1}) & \text{if} ~~N_{\text{jets}}\geq 1 \\
      0 & \text{if} ~~N_{\text{jets}}=0 
   \end{cases}~~,
\end{equation}
where $p_T^{\rm jet_1}$ is the $p_T$ of the leading jet. 

\item The di-jet transverse mass $m^{\rm di-jet}_T$ of the leading and sub-leading jets. It is a good discriminant against QCD multijet which tends to have a large value of this variable. If two jets cannot be found in an event, this variable is set to zero.

\item The effective mass defined as
\begin{equation}
m_{\rm eff}=H_T+E^{\rm miss}_T+p_T(\tau_{h1})+p_T(\tau_{h2}),
\end{equation}
where $H_T$ is the sum of the hadronic $p_T$'s in an event, $p_T(\tau_{h1})$ and $p_T(\tau_{h2})$ are the transverse momenta of the leading and sub-leading hadronic taus. 

\end{enumerate}

BDTs employ a multivariate analysis technique to a classification problem such as the one at hand. The aim here is to classify a certain set of events as belonging to the signal or the background by using a number of discriminating variables (1$-$8 listed above) to make the decision. The signal $(S)$ and background $(B)$ samples undergo two phases: the training and testing phases. In the first phase, the BDTs are trained on those samples using the available list of kinematic variables. The algorithm sorts those variables in a descending order of separation power and chooses the variable that has the highest separation power to start the ``root node". A cut is applied on this variable and events are split into left or right nodes depending whether they were classified as signal or background. Afterwards, another variable (or sometimes the initiating variable is kept) is chosen with a cut value which further splits down events into signal and background. The tree continues growing until a stopping criterion, such as the tree depth, is reached. The end layer of the tree contains the leaves which host the events classified as signal and given a value $+1$ or background and given a value $-1$. During training, some signal events may be misclassified as background and vice-versa. Those events will be given a weight factor and then enter in the second iteration of the training phase when the next tree starts forming. Those events are now given more attention thanks to the weight factor they carry. The training stops when the entire number of trees in the forest have been utilized. The number of trees and their depths are specified by the user in such a way as to maximize the separation between the signal and the background where a larger depth generally produces a better separation.

 The second phase is the testing phase where the algorithm applies what it has learnt on a statistically independent set of samples and outputs a new discriminating variable called the ``BDT score" or ``BDT response". An agreement between the performances of the training and testing phases is a sign of no overtraining occurring in the analysis. Such a situation can arise if one chooses too large of a tree depth while not having enough statistics in the samples. We have made sure that no overtraining of the samples has occurred throughout. The BDT implementation is carried out
using \code{ROOT}'s own TMVA (Toolkit for Multivariate Analysis) framework~\cite{Speckmayer:2010zz}. Depending on the samples, we set the number of trees to be in the range 120 to 200, the depth to 3 and the \code{AdaBoost} learning rate to 0.5. Many combinations of those parameters have been tried and the one which gave the best result was considered. \\
BDTs are very useful in classification problems where conventional linear cuts fail. To show that, we display in Fig.~\ref{fig2} distributions normalized to unity of four kinematic variables at $\sqrt{s}=27$ TeV for benchmark (a) of Table~\ref{tab1}. The purpose of such distributions is to help design event selection cuts which would allow better background rejection based on the shape of the distribution. One can clearly notice, for distributions in $\ln(p^{\rm jet}_T)$, $N^{\tau}_{\rm tracks}$ and $\Delta\phi(\tau_{h1},\tau_{h2})$, a conventional linear cut does not do the job as it would lead to a poor signal to background ratio. This is where BDTs become powerful since they employ non-linear cuts in the multidimensional space of variables (thus the name multivariate analysis). On the other hand, a linear cut on $m^{\rm tot}_T$ such that $m^{\rm tot}_T>150$ GeV is reasonable but not sufficient to obtain a good signal to background ratio. The BDT algorithm will run through distributions of such sort for the eight variables presented above and design unconventional cuts to obtain the best discrimination between the signal and the background. 

\begin{figure}[H]
 \centering
 	\includegraphics[width=0.4\textwidth]{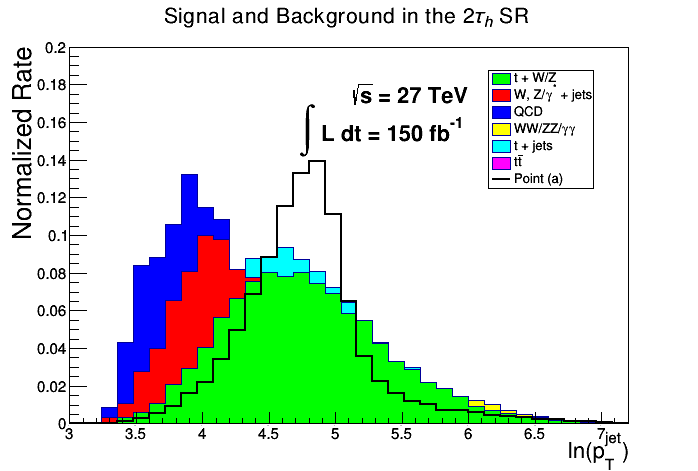}
 	\includegraphics[width=0.4\textwidth]{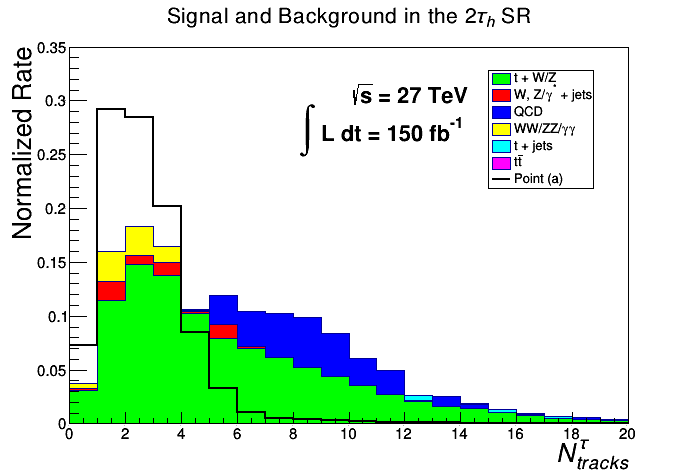} \\
 	\includegraphics[width=0.4\textwidth]{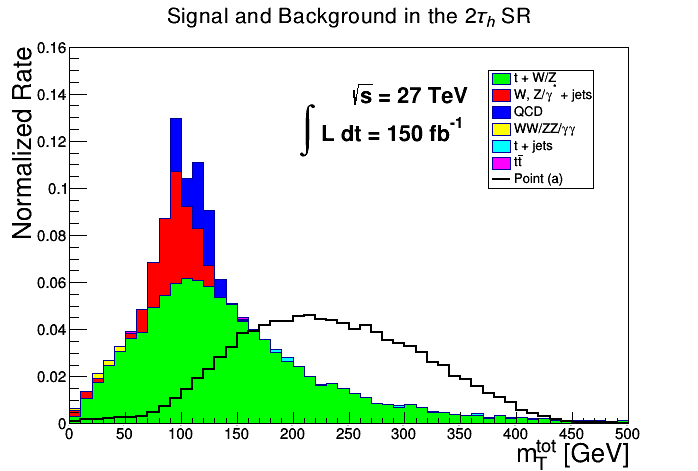}
 	\includegraphics[width=0.4\textwidth]{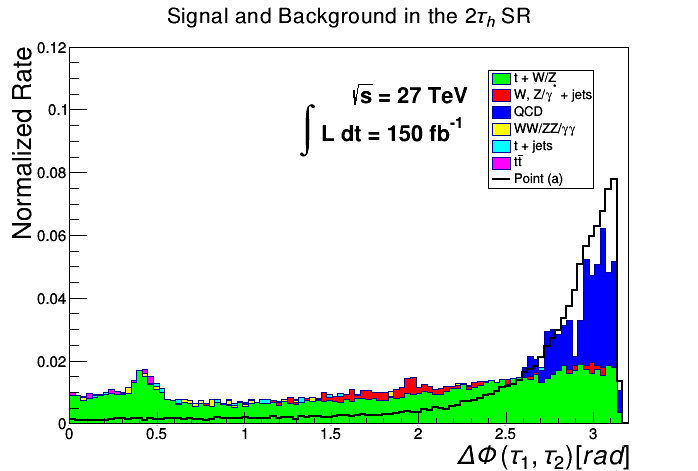}
      \caption{Distributions normalized to the bin size of four kinematic variables for benchmark (a) at 27 TeV: $\ln(p^{\rm jet}_T)$ (top left), $N^{\tau}_{\rm tracks}$ (top right), $m^{\rm tot}_T$ (bottom left) and $\Delta\phi(\tau_{h1},\tau_{h2})$ (bottom right) in the $2\tau_h$ signal region (SR).  }
	\label{fig2}
\end{figure}

After training and testing of the BDTs, we set a cut on the BDT score variable which would give us the minimum integrated luminosity for $\frac{S}{\sqrt{S+B}}$ at the $5\sigma$ level discovery. In general, this cut value is not common across all points since each point is trained and tested separately along with the SM backgrounds and so the distribution in BDT score differs from one point to another. We present in Fig.~\ref{fig3} the computed 
 integrated luminosities, $\mathcal{L}$, as a function of the cut on the BDT score for both 14 TeV (left panel) and 27 TeV (right panel) machines. For 14 TeV, one can see that a drop in $\mathcal{L}$ occurs for BDT score $>0.3$ while at 27 TeV the same is seen for BDT score $>0.2$.   

\begin{figure}[H]
 \centering
 	\includegraphics[width=0.48\textwidth]{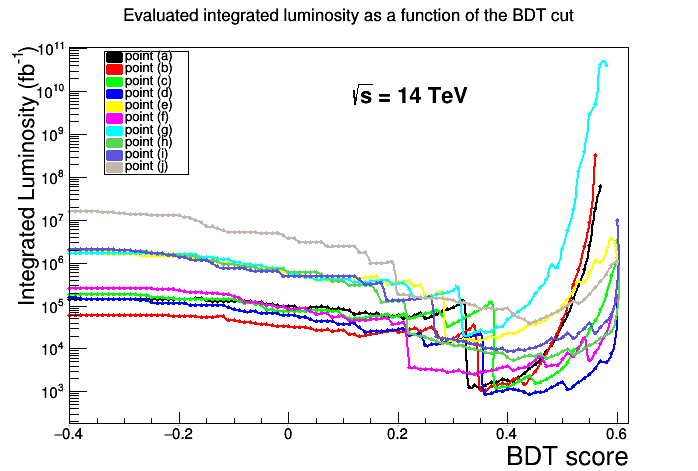}
 	\includegraphics[width=0.48\textwidth]{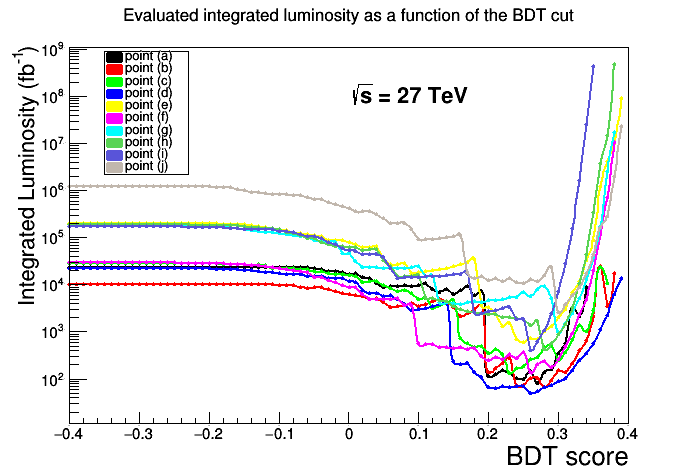} 
      \caption{The estimated      
       integrated luminosities as a function of the BDT cut for the benchmarks of Table~\ref{tab1} at 14 TeV (left panel) and 27 TeV (right panel).}
	\label{fig3}
\end{figure}

Using the results from Fig.~\ref{fig3}, we tabulate the lowest integrated luminosity for a $5\sigma$ discovery at HL-LHC and HE-LHC in Table~\ref{tab7} for all our benchmarks. It is seen that half the benchmarks are discoverable at HL-LHC.
Thus benchmark (d) is discoverable  
 with an $\mathcal{L}$ as low as 866$\ifb$ while the benchmark (f)  requires  $\mathcal{L}$ close to the optimal integrated luminosity of 3000$\ifb$. However, all  the benchmarks are discoverable at HE-LHC with some requiring an integrated luminosity smaller than 100$\ifb$ 
 such as point (d) with $\mathcal{L}=50\ifb$ for discovery. Point (j) requires the largest amount of data at $\sim 2600\ifb$ 
 which, however,  is still much lower than the optimal integrated luminosity of 15 ab$^{-1}$ expected at HE-LHC.

\begin{table}[H]
	\centering
	\begin{tabulary}{\linewidth}{l|c|c}
    \hline\hline
    & \multicolumn{2}{c}{$\mathcal{L}$ for $5\sigma$ discovery in 2$\tau_h$ + b-jets}  \\
	\hline
	Model & $\mathcal{L}$ at 14 TeV & $\mathcal{L}$ at 27 TeV \\
	\hline
  (a) & 1221 & 82 \\
  (b) & 1102 & 67 \\ 
  (c) & 1195 & 131 \\
  (d) & 866 & 50 \\ 
  (e) & ... & 604 \\ 
  (f) & 2598 & 136 \\ 
  (g) & ... & 952 \\ 
  (h) & ... & 420 \\ 
  (i) & ... & 412 \\ 
  (j) & ... & 2599 \\ 
	\hline
	\end{tabulary}
	\caption{Comparison between the estimated integrated luminosity ($\mathcal{L}$) in fb$^{-1}$ 
	for a 5$\sigma$ discovery at 14 TeV (middle column)
	 and 27 TeV (right column)
		 for the CP odd Higgs following the selection criteria and BDT cut.
		 An entry with an ellipsis means that the evaluated $\mathcal{L}$ is much greater than $3000\ifb$.}
\label{tab7}
\end{table}

We show in Fig.~\ref{fig4} some distributions in the BDT score for points (a) and (d) at 14 TeV and 27 TeV for various representative integrated luminosities. For point (a) which is discoverable at both HL-LHC and HE-LHC we  see that the signal is in excess over the background for $\mathcal{L}=150 \ifb$ at 27 TeV (top left panel) while higher integrated luminosity is required for an excess at HL-LHC, namely,  $\mathcal{L}=2000 \ifb$ (top right panel). Note that those integrated luminosities are not the minimum ones with the latter presented in Table~\ref{tab7}. The bottom two panels of Fig.~\ref{fig4} show the BDT score for point (d) at 27 TeV (left) and 14 TeV (right) for the same integrated luminosity of $200\ifb$. Notice that the signal is in excess over the background at HE-LHC contrary to the HL-LHC case which is in accordance with the data of Table~\ref{tab7}.  

\begin{figure}[H]
 \centering
 	\includegraphics[width=0.45\textwidth]{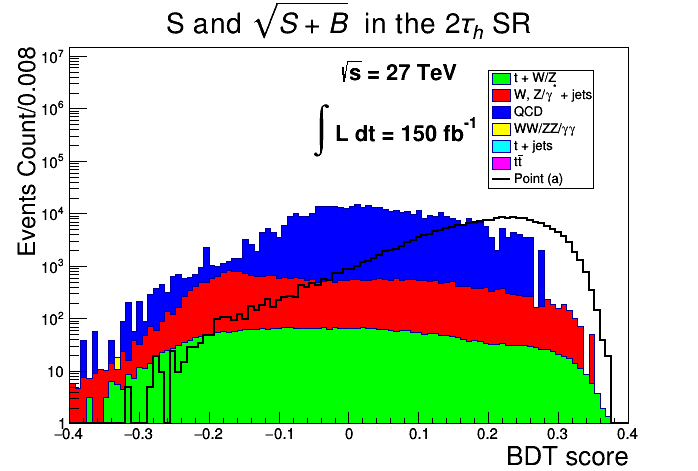}
 	\includegraphics[width=0.45\textwidth]{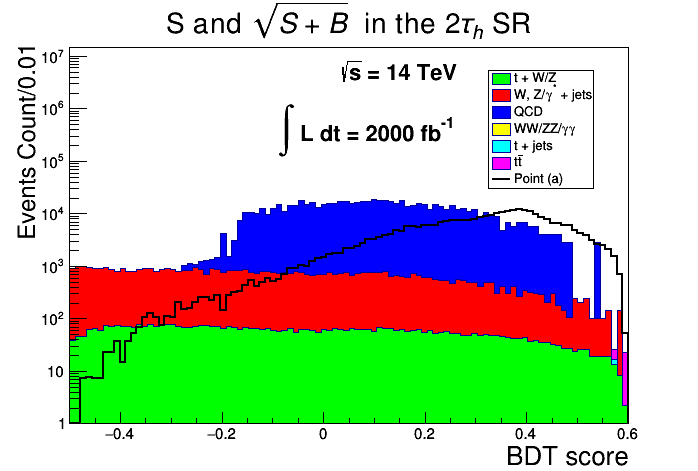}\\
 	\includegraphics[width=0.45\textwidth]{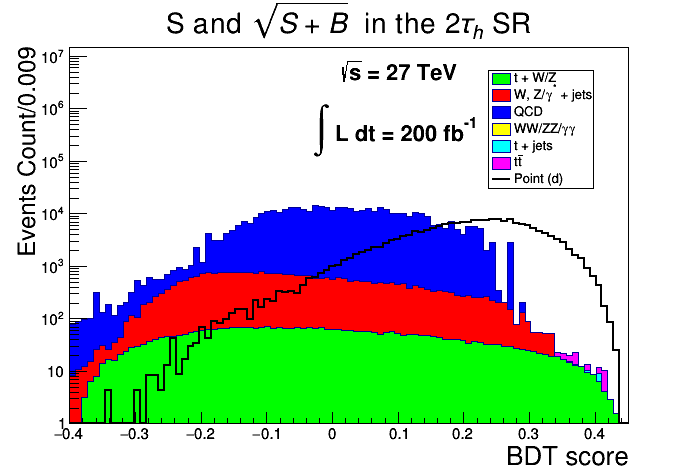}
 	\includegraphics[width=0.45\textwidth]{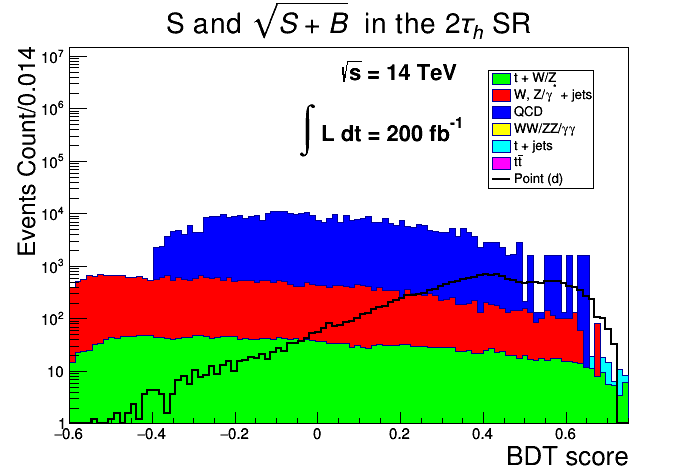}
      \caption{Distributions in the BDT score for benchmarks (a) (top panels) and (d) (bottom panels) of Table~\ref{tab1} at 14 TeV (right panels) and 27 TeV (left panels) in the $2\tau_h$ signal region (SR) for various integrated luminosities. }
	\label{fig4}
\end{figure} 

It is worth mentioning that in using BDTs, it is seen that better separation between signal and background occurs for points with higher CP odd Higgs masses due to more energetic final states. However, a better outcome, i.e. a smaller integrated luminosity, is not always seen in those cases since the price to be paid for larger masses is  a falling cross-section which results in much higher integrated luminosity for discovery. Here we investigate the possibility if at higher masses  the cross-section, $\sigma\times \text{BR}(A\rightarrow \tau\tau)$, can be maintained at a larger than usual value. This can be achieved for a higher branching ratio and larger $\tan\beta$. Benchmark (b) has the largest cross-section amongst all the points but requires the second lowest integrated luminosity for discovery with point (d) requiring the least. Now point (d) has a CP odd Higgs mass 100 GeV heavier than point (b) but has a higher $\tan\beta$ and branching ratio which makes up for the mass increase and keeps the cross-section from falling too rapidly. Because of the effect of $\tan\beta$, the branching ratio and more energetic final states, the separation between signal and background for point (d) is more pronounced leading to the lowest integrated luminosity for discovery. 

Given that the HE-LHC is expected to collect data at the rate of 820$\ifb$ per year~\cite{HE-LHC-1} versus the rate at which HL-LHC will collect data, the projected runtime for points (a)$-$(d) for discovery at HL-LHC is $\sim 3$ to $\sim 4$ years while point (f) requires $\sim 8$ years. The runtime is significantly decreased for HE-LHC where most of the points require $\sim 1$ to $\sim 6$ months, while point (d) $\sim 22$ days and point (j) $\sim 3$ years.   

Before concluding we give an overview of the uncertainties one might expect and their impact on the estimated integrated luminosities at HL-LHC and HE-LHC. One of the main challenges to overcome in experiment while analyzing data are the systematic uncertainties. One would expect such uncertainties to decrease when HL-LHC starts operation due to an increased data set. It is also reasonable to assume that improvements on this front is expected by experimentalists working on ATLAS and CMS detectors. Following experts' opinion in the HL-LHC and HE-LHC Working Groups at CERN~\cite{Cepeda:2019klc,CidVidal:2018eel}, much of the systematic uncertainties are expected to drop by a factor of 2. Those uncertainties are known as ``YR18 systematic uncertainties". In the $A\rightarrow\tau_h\tau_h$ channel, systematic uncertainties due to the estimation of QCD jets to $\tau_h$ fake background are dominant especially in the low CP odd Higgs mass region. For higher masses, the leading uncertainty is from the reconstruction and
identification of high transverse momentum $\tau_h$. In this analysis we assume that the systematic uncertainties in the background and signal are $20\%$ and $10\%$, respectively. We give higher (lower) combined uncertainty for points with smaller (larger) Higgs mass. The integrated luminosity for a $5\sigma$ discovery is re-estimated after including the uncertainties using the signal significance~\cite{Cowan:2010js}
\begin{equation}
\sigma=\left[2\left((S+B)\ln\left[\frac{(S+B)(B+\Delta_C^2)}{B^2+(S+B)\Delta^2_C}\right]-\frac{B^2}{\Delta^2_C}\ln\left[1+\frac{\Delta^2_C S}{B(B+\Delta^2_C)}\right]\right)\right]^{1/2},
\end{equation}
where $\Delta_C$ is the combined uncertainty in signal and background, $\Delta^2_C=\Delta^2_S+\Delta^2_B$. We show in Fig.~\ref{fig5} the estimated integrated luminosities before and after including the uncertainties for both HL-LHC and HE-LHC. In the left panel, the five benchmarks discoverable at both machines are shown along with the ``YR18 uncertainties" where at HL-LHC, the integrated luminosities have increased by $\sim 1.5$  to $\sim 2.5$ times (in blue) compared to when no systematic uncertainties are present (in orange). At the HE-LHC the increase is by $\sim 1.5$ to $\sim 4$ times (in red) compared to the case with no systematics (in yellow). The right panel shows the points that are discoverable only at HE-LHC along with the integrated luminosities before (in orange) and after (in blue) including uncertainties.

\begin{figure}[H]
 \centering
 	\includegraphics[width=0.49\textwidth]{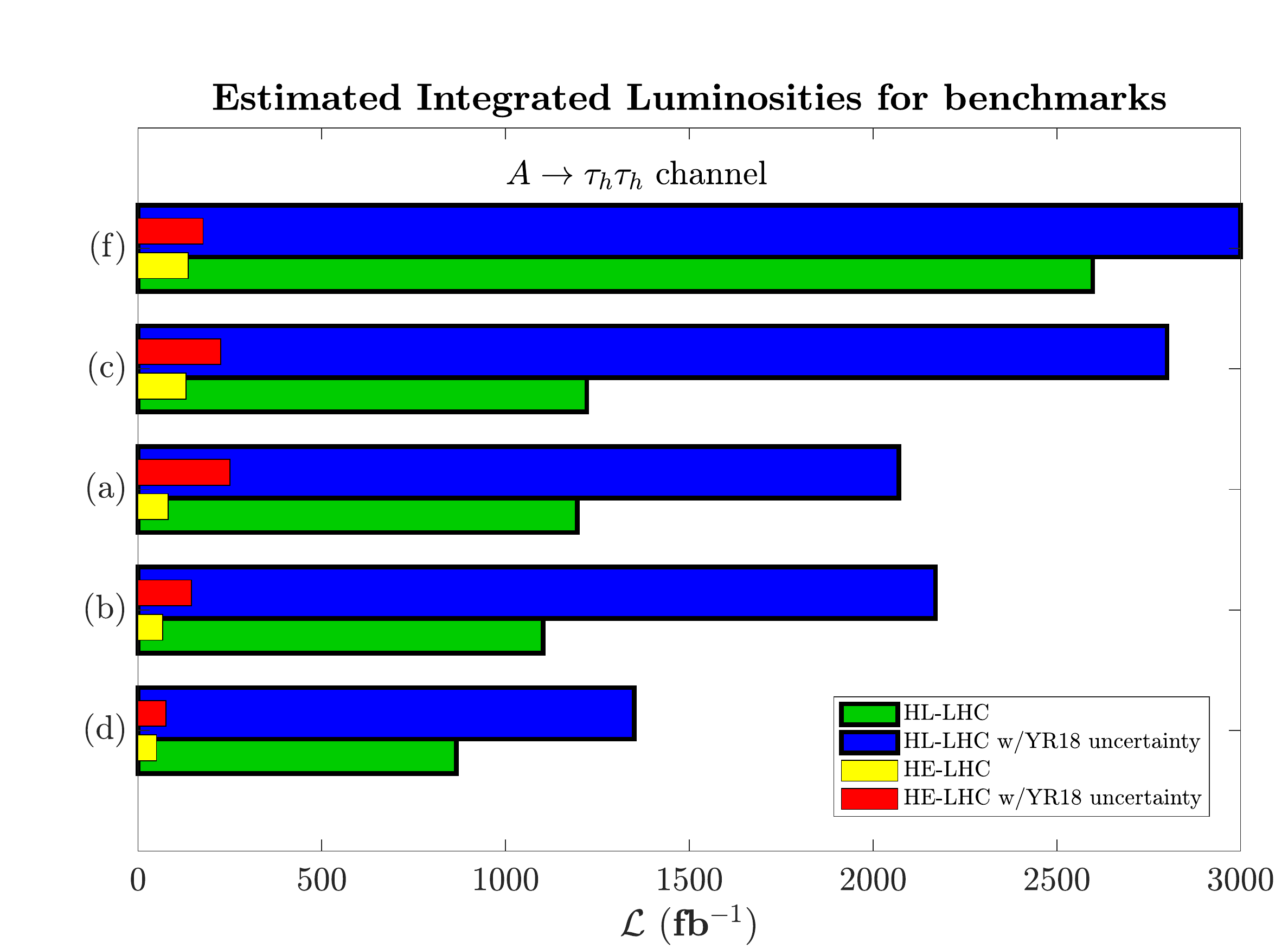}
 	\includegraphics[width=0.49\textwidth]{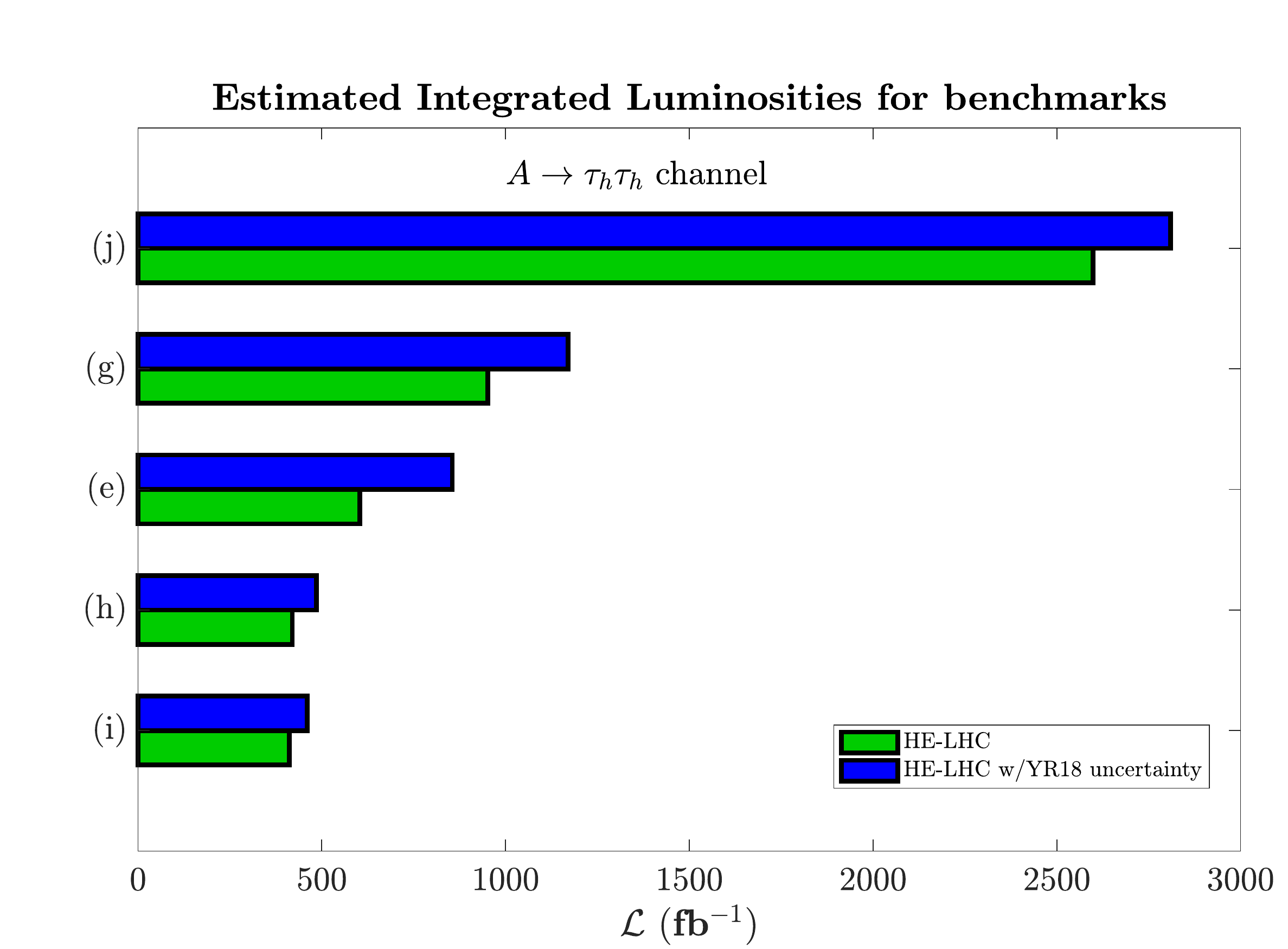}
      \caption{Left panel:  five benchmarks of Table~\ref{tab1} that are discoverable at both HL-LHC and HE-LHC 
      with and without       
       the ``YR18 uncertainties". Right panel: the  remaining five benchmarks of Table~\ref{tab1}       
        that are discoverable only at HE-LHC  with  and without 
             the ``YR18 uncertainties".}
	\label{fig5}
\end{figure}

\section{Conclusions}\label{sec:conc}

The large size of weak scale supersymmetry implied by the Higgs boson mass at $\sim 125$ GeV has a direct implication for 
 the discovery of supersymmetric dark matter. Thus typically in high scale SUGRA models with a large size of weak scale supersymmetry
 often the LSP neutralino is mostly a bino making an efficient annihilation of bino dark matter in the early universe difficult
 and
 leading to its overabundance inconsistent with observation. Often coannihilation is utilized in this case where
 a low-lying next-to-LSP and the LSP together act to reduce the relic density within the observed limit.  However, another branch of 
 radiative breaking of the electroweak symmetry exists within the SUGRA model where a large size of weak scale supersymmetry 
 can coexist with a small Higgs mixing parameter $\mu$ (of size the electroweak scale).  Such a situation can lead to 
 a higgsino-like LSP. As mentioned in the introduction, 
  models of this type are severely constrained by simultaneous satisfaction of dark matter relic density and by the 
 spin-independent proton-DM scattering cross section  limits in  direct detection experiments. 
 However, such models can be made viable if the dark matter is multi-component. 
  Thus in this work we considered an 
  extension of the MSSM/SUGRA gauge group with a  $U(1)_X$  of the hidden sector where the $U(1)_X$  
  and $U(1)_Y$ have kinetic and Stueckelberg mass mixings. Further, we assume that the hidden sector is populated with chiral
  matter leading to a Dirac fermion which acts as the second component of dark matter and makes up the dark matter deficit
  in the relic density. One implication of the model is the existence of a relatively light CP odd Higgs boson $A$, as well as 
  relatively light $H$ and $H^{\pm}$, which have masses of the electroweak size. We investigate a set of benchmarks 
  for the extended model and show that the CP odd Higgs boson in models of this type is observable in the next generation 
  collider experiments. Specifically we investigate the discovery potential of a CP odd Higgs in the $\tau_h\tau_h$ final state at the HL-LHC and HE-LHC. It is seen that a CP odd Higgs with a mass up to 450 GeV for $\tan\beta\leq 12$ may be discoverable at HL-LHC. The discovery reaches 750 GeV at the HE-LHC with an integrated luminosity of $\sim 2600$ fb$^{-1}$ which is just a fraction of the optimal luminosity of 
  15 ab$^{-1}$ that HE-LHC can deliver. With the optimal luminosity the mass reach of HE-LHC for the CP odd Higgs mass will
  certainly extend far above 750 GeV. It is also shown that a significant part of the parameter space of the extended model
  can be probed in the next generation direct detection experiments such as XENONnT and LUX-ZEPLIN.

\vspace{2cm}

\textbf{Acknowledgments: }
The analysis presented here was done using the resources of the high-performance  Cluster353 at the Advanced Scientific Computing Initiative (ASCI) and the Discovery Cluster at Northeastern University.  This research was supported in part by the NSF Grant PHY-1620575.

\end{document}